\documentclass[twoside,10pt]{article}
\usepackage{CJK}
\usepackage{indentfirst}
\usepackage{bm}
\usepackage[]{hyperref}
\usepackage[numbers,sort&compress]{natbib}
\usepackage{tikz}
\usepackage{float}
\usepackage{color}
\usepackage{graphicx}
\usepackage{subfig}
\usepackage{amsmath}
\usepackage{mathrsfs}
\usepackage{amssymb}
\usepackage{amsthm}
\usepackage{epstopdf}
\newcommand{\RNum}[1]{\uppercase\expandafter{\romannumeral #1\relax}}

\makeatletter
\@addtoreset{equation}{section}
\makeatother
\begin{document}
\begin{CJK*}{GBK}{song}
\allowdisplaybreaks


\begin{center}
\LARGE\bf  The general solutions for a non-isospectral integrable TD hierarchy via the inverse scattering transform
\end{center}
\footnotetext{\hspace*{-.45cm}\footnotesize $^*$Corresponding author: B.L. Feng. E-mail: zyfusf@163.com }
\begin{center}
\ \\Hongyi Zhang$^{1}$, Yufeng Zhang$^{1}$, Binlu Feng$^{2*}$
\end{center}
\begin{center}
\begin{small} \sl
{$^{1}$School of Mathematics, China University of Mining and Technology, Xuzhou, Jiangsu, 221116, People's Republic of China.\\
$^{2}$School of Mathematics and Information Sciences, Weifang University, Weifang,Shandong, 261061, People's Republic of China.}
\end{small}
\end{center}
\vspace*{2mm}
\begin{center}
\begin{minipage}{15.5cm}
\parindent 20pt\footnotesize


 \noindent {\bfseries
Abstract}
A non-isospectral Lax pair is first introduced from which a kind of non-isospectral integrable TD hierarchy is derived, whose reduction is an integrable system called the non-isospectral integrable TD system. Then by using
the inverse scattering transform (IST) method, new general soliton solutions for the non-isospectral integrable TD hierarchy are obtained. Because we investigate soliton solutions of non-isospectral integrable systems by the IST method, a new Gel'fand-Levitan-Marchenko (GLM) equation needs to be constructed. Finally, we explicitly obtain the exact solutions of the non-isospectral integrable TD system. The method presented in the paper can be extensively applied to other integrable equations.

\end{minipage}
\end{center}
\begin{center}
\begin{minipage}{15.5cm}
\begin{minipage}[t]{2.3cm}{\bf Keywords}\end{minipage}
\begin{minipage}[t]{13.1cm}
Exact solution; Non-isospectral TD hierarchy; Inverse scattering transform
\end{minipage}\par\vglue8pt
\end{minipage}
\end{center}

\section{Introduction}  
The inverse scattering transform (IST) was initially proposed by Gardner et al. in 1967 \cite{Gardner67}, who successfully applied them to the Korteweg-de Vries (KdV) equation.
The classical IST methods are generally studied using the Gel'fand-Levitan-Marchenko integral equation as a foundation.
In 1984, Zakharov et al. appropriately simplified the IST method by issuing the Riemann-Hilbert (RH) formula \cite{Zakharov84}.
Subsequently, more and more researchers have utilized this method to obtain the soliton solutions for nonlinear evolution equations \cite{Ablowitz81,Ablowitz91,Ji17,Ma19,Ning04,Ning042,Li08,Li10,Li12}.
In general, there are two kinds of evolution equations associated with the same spectral problems, called isospectral hierarchy and non-isospectral hierarchy, respectively.
  The isospectral equations mainly describe solitary waves in lossless homogeneous media.
These equations describe situations where the spectral parameters remain constant with time.
  In recent years, the derivation, reduction and application of non-isospectral equations have been well developed \cite{Zhang23,Wang20,Wang202,Wang203}.
In contrast, the non-isospectral equations arise from the spectral problem with time-dependent spectral parameter $\lambda$ \cite{Calogero78,Calogero82,Li82,Ma92,Ma922,Chen96}, and IST can be efficiently applied to solve these equations, and their solutions prove the existence of solitary waves in certain types of inhomogeneous media \cite{Hirota76,Gupta79,Chan89}.
 In addition, the search for exact solutions to nonisospectral equations is of great importance mathematically due to the involvement of time-dependent spectral parameters. Previous discussions have addressed some issues related to seeking exact solutions for nonisospectral equations.
 Additionally, some non-isospectral equations have been successfully solved by using IST method \cite{Gupta79,Chan89,Zhang22}.

The TD hierarchy associated with spectral problem
\begin{equation}\label{sp1}
\Psi_{x}=\left(\begin{array}{cc}
-\lambda+s & q \\
r & \lambda
\end{array}\right)\Psi, \quad \Psi=(\Psi_{1},\Psi_{2})^T,
\end{equation}
is one of the integrable hierarchies series (ranging from the TA hierarchy to the TD hierarchy) proposed and named by Tu \cite{Tu891,Tu892}. Spectrum problem \eqref{sp1} is equivalently to spectral problem \cite{Cao93}
\begin{equation}\label{sp2}
\Psi_{x}=\left(\begin{array}{cc}
i\lambda+u & v \\
v & -i\lambda-u
\end{array}\right)\Psi, \quad \Psi=(\Psi_{1},\Psi_{2})^T.
\end{equation}
Zhu \cite{Zhu18,Zhu19} utilized \eqref{sp2} to get the so-called TD equation:
\begin{equation}\label{td}
\begin{cases}
iu_{t}+\left(\frac{v_{xx}}{4v}+u^2-\frac{v^2}{2}\right)_x=0,\\
iv_{t}+u_xv+2uv_x=0.
\end{cases}
\end{equation}
In the paper, by choosing new expressions for the time part of the Lax pair, through a non-isospectral zero-curvature equation, a new non-isospectral integrable TD system (see below \eqref{equ2}) is worked out, meanwhile, the TD equation \eqref{td} is only its special case in isospectral case.
In Section 2, we depict the derivation of the non-isospectral TD hierarchy.
In Section 3, we obtain the general solutions of the non-isospectral integrable TD hierarchy through the IST method. Based on our paper \cite{zhang022}, we analyze dynamic behaviors of the soliton solutions.

\section{A non-isospectral integrable TD hierarchy}  
Consider the TD spectral problem \cite{Cao93}
\begin{equation}\label{lax1}
\Psi_{x}=M \Psi, \quad M=\left(\begin{array}{cc}
i\lambda+u & v \\
v & -i\lambda-u
\end{array}\right), \quad \Psi=(\Psi_{1},\Psi_{2})
\end{equation}
and the time evolution introduced by us:
\begin{equation}
\Psi_{t}=N \Psi, \quad N=\left(\begin{array}{cc}
A & B+C \\
B-C & -A
\end{array}\right),
\end{equation}
where $u(x,t)$ and $v(x,t)$ are potential functions, $\lambda$ is a spectral parameter, and $A=\sum_{i=0}^{n+2}A_i\lambda^{n+2-i}$, $B=\sum_{i=0}^{n+1}B_i\lambda^{n+1-i}$, $C=\sum_{i=0}^{n}C_i\lambda^{n-i}$. We assume that $u(x,t)$ and $v(x,t)$ are smooth functions of $x$ and $t$; and their derivatives of any order with respect to $x$ vanish rapidly as $x \rightarrow \infty $. The compatibility condition reads
\begin{equation}\label{zero1}
M_{t}-N_{x}+[M, N]=0,
\end{equation}
expanding the matrices $M$ and $N$ gives
\begin{equation}\label{a1}
\begin{cases}
i\lambda_{t}+u_{t}-A_x-2vC=0,\\
v_t-B_x-C_{x}-2vA+2(i\lambda+u)(B+C)=0,\\
v_t-B_x+C_{x}+2vA-2(i\lambda+u)(B-C)=0.
\end{cases}
\end{equation}
Simplifying the above equation, we get
\begin{align}
&i\lambda_{t}+u_{t}-A_x-2vC=0, \label{a11}\\
&v_t-B_x+2(i\lambda+u)C=0, \label{a12}\\
&C_{x}+2vA-2(i\lambda+u)B=0. \label{a13}
\end{align}
Comparing the coefficients of $\lambda^{l}(l=0,1,\cdots n+1)$ in \eqref{a11} and \eqref{a12} respectively, we get
\begin{equation} \label{a14}
\begin{cases}
u_{t}-A_{n+2,x}-2vC_{n}=0, \\
-A_{i+2,x}-2vC_{i}=0, \\
i\lambda_{t}-A_{1,x}\lambda^{n+1}-A_{0,x}\lambda^{n+2}=0,\\
v_{t}-B_{n+1,x}+2uC_{n}=0, \\
-B_{i+1,x}+2uC_{i}+2iC_{i+1}=0, \\
-B_{0,x}+2iC_{0}=0.
\end{cases}
\end{equation}
Then we consider the coefficients of $\lambda^{l}(l=0,1,\cdots n+2)$ in \eqref{a13}, to get
\begin{equation} \label{a15}
\begin{cases}
C_{n,x}+2vA_{n+2}-2uB_{n+1}=0,\\
C_{i,x}+2vA_{i+2}-2iB_{i+2}-2uB_{i+1}=0,\\
2vA_{1}-2iB_{1}-2uB_{0}=0,\\
2vA_{0}-2iB_{0}=0.
\end{cases}
\end{equation}
In contrast to the isospectral case, the nonisospectral case involves spectral parameters that depend on time \cite{Calogero78,Calogero82,Li82,Ma92,Ma922,Chen96}. We choose $\lambda_{t}=\lambda^{n+1}$ here, and take the initial values\\
$~~~~~~~~~~~~~~~~~~~~~~~~~~~~~~~~~~~~~~~~~~~~~~~~~~~~~~~~A_0=0$,~~$B_0=0$,~~$C_0=0$,\\
which can be obtained by equations \eqref{a14} and \eqref{a15}\\
$~~~~~~~~~~~~~~~~~~~~~~~~~~~~~~~~~~~~~~~A_1=ix$,~~$B_1=xv$,~~$C_1=-\frac{i}{2}(v+xv_{x})$,~~$A_{2}=0$,~~$B_{2}=ixuv$.\\
For $i=0,\cdots n-1$, from equations \eqref{a14} and \eqref{a15} we can get
\begin{equation} \label{a16}
\begin{cases}
-A_{i+2,x}-2vC_{i}=0,\\
-B_{i+1,x}+2uC_{i}+2iC_{i+1}=0,\\
C_{i,x}+2vA_{i+2}-2iB_{i+2}-2uB_{i+1}=0.
\end{cases}
\end{equation}
From equation \eqref{a16}, we get the following recurrence relation
\begin{equation}
\left(\begin{array}{c}
B_{i+2} \\
C_{i+1}
\end{array}\right)=
\left(\begin{array}{cc}
iu & iv\partial^{-1}(2v)-\frac{i}{2}\partial \\
-\frac{i}{2}\partial & iu
\end{array}\right)\left(\begin{array}{c}
B_{i+1} \\
C_i
\end{array}\right)=L\left(\begin{array}{c}
B_{i+1} \\
C_i
\end{array}\right),
\end{equation}
where\\
$~~~~~~~~~~~~~~~~~~~~~~~~~~~~~~~~~~~~~~~~~~~~~~~L=\left(\begin{array}{cc}
iu & iv\partial^{-1}(2v)-\frac{i}{2}\partial \\
-\frac{i}{2}\partial & iu
\end{array}\right)$.\\
From the values of $B_i$ and $C_i$ we can get the value of $A_i$. For $n=0$, we get $A_{1}=ix$, $A_{2}=xu$ by \eqref{a15}; For $n\geq 1$,
by equations \eqref{a14} and \eqref{a15} we obtain $A_{i+2}=\partial^{-1}(-2vC_{i})$ $(i=0,1,\cdots n-1)$ and  $A_{n+2}=\frac{2uB_{n+1}-C_{n,x}}{2v}$ respectively. Finally, we obtain the TD non-isospectral equation
\begin{equation}
\begin{cases}
u_{t}=2vC_{n}+\left(\frac{2uB_{n+1}-C_{n,x}}{2v}\right)_{x},\\
v_{t}=B_{n+1,x}-2uC_{n},\label{equ1}
\end{cases}
\end{equation}
when $n=1$, \eqref{equ1} reduced to:
\begin{equation} \label{equ2}
\begin{cases}
u_{t}=-i\left(v^2+xvv_{x}\right)+\left(\frac{2iu^2vx+\frac{i}{2}\left(
v+xv_{x}\right)_{x}}{2v}\right)_{x},\\
v_{t}=B_{2,x}-2uC_{1}=2iu\left(v+xv_{x}\right)+ixu_{x}v.
\end{cases}
\end{equation}

$\mathbf{Remark ~2.1.}$ Setting $\lambda_t=0$ in \eqref{lax1}, by the same method as in the previous non-isospectral equations, we get
\begin{equation}
\begin{cases}
u_{t}=2vC_{n}+\left(\frac{2uB_{n+1}-C_{n,x}}{2v}\right)_{x},\\
v_{t}=B_{n+1,x}-2uC_{n}.\label{equ4}
\end{cases}
\end{equation}
with
\begin{equation}
\begin{cases}
B_1=uvA_0-ivA_1, C_0=-\frac{1}{2}A_0v_x,\\
\left(\begin{array}{c}
B_{i+2} \\
C_{i+1}
\end{array}\right)=L\left(\begin{array}{c}
B_{i+1} \\
C_i
\end{array}\right),~(i=0,\cdots,n)\label{equ222}
\end{cases}
\end{equation}
where $A_0, A_1$ are arbitrary constants. In this case, the isospectral hierarchy \eqref{equ4} is equivalent to the isospectral TD hierarchy \cite{Tu892,Cao93}.

Moreover, we note that \eqref{equ4} becomes the TD equation \eqref{td} when $A_0=0, A_1=i$ and $n=1$.

\section{ General solution of the non-isospectral TD hierarchy}
In this section, we will give the direct scattering problem and the time evolution of the scattering data, which holds for the entire hierarchy.
Finally, by using the classical inverse scattering transform, we obtain the N-soliton solution of the entire non-isospectral TD hierarchy.

\subsection{The direct scattering problem}
$\mathbf{Lemma~ 3.1.}$ If potential $(u(x), v(x))^{T}$ satisfies
\begin{equation}
\int_{-\infty}^{\infty}|x^{j}u(x)|dx<+\infty,~~\int_{-\infty}^{\infty}|x^{j}v(x)|dx<+\infty~~(j=0,1),
\end{equation}
then the spectral problem \eqref{lax1} has a group of Jost solutions $\phi(x;\lambda)$, $\widetilde{\phi}(x;\lambda)$, $\psi(x;\lambda)$ and $\widetilde{\psi}(x;\lambda)$ which are bounded for all values of $x$, and also have the following asymptotic behaviors:\\
 when $x\rightarrow +\infty$,
\begin{equation}\label{asym1}
\begin{array}{l}
\phi(x;\lambda)=\left(\begin{array}{c}
\phi_1 \\
\phi_2
\end{array}\right)\rightarrow \left(\begin{array}{c}
1 \\
0
\end{array}\right)e^{i\lambda x},~~~
\widetilde{\phi}(x;\lambda)=\left(\begin{array}{c}
\widetilde{\phi}_1 \\
\widetilde{\phi}_2
\end{array}\right)\rightarrow \left(\begin{array}{c}
0 \\
1
\end{array}\right)e^{-i\lambda x},\\
\phi_{x}(x;\lambda)\rightarrow i\lambda \left(\begin{array}{c}
1 \\
0
\end{array}\right)e^{i\lambda x},~~~
\widetilde{\phi}_{x}(x;\lambda)\rightarrow -i\lambda\left(\begin{array}{c}
0 \\
1
\end{array}\right)e^{-i\lambda x},\\
\phi_{\lambda}(x;\lambda)\rightarrow ix \left(\begin{array}{c}
1 \\
0
\end{array}\right)e^{i\lambda x},~~~
\widetilde{\phi}_{\lambda}(x;\lambda)\rightarrow -ix\left(\begin{array}{c}
0 \\
1
\end{array}\right)e^{-i\lambda x},
\end{array}
\end{equation}
and when $x \rightarrow -\infty$,
\begin{equation}\label{asym2}
\begin{array}{l}
\psi(x;\lambda)=\left(\begin{array}{c}
\psi_1 \\
\psi_2
\end{array}\right)\rightarrow \left(\begin{array}{c}
0 \\
1
\end{array}\right)e^{-i\lambda x},~~~
\widetilde{\psi}(x;\lambda)=\left(\begin{array}{c}
\widetilde{\psi}_1 \\
\widetilde{\psi}_2
\end{array}\right)\rightarrow \left(\begin{array}{c}
1 \\
0
\end{array}\right)e^{i\lambda x},\\
\psi_{x}(x;\lambda)\rightarrow -i\lambda \left(\begin{array}{c}
0 \\
1
\end{array}\right)e^{-i\lambda x},~~~
\widetilde{\psi}_{x}(x;\lambda)\rightarrow i\lambda \left(\begin{array}{c}
1 \\
0
\end{array}\right)e^{i\lambda x},\\
\psi_{\lambda}(x;\lambda)\rightarrow -ix \left(\begin{array}{c}
0 \\
1
\end{array}\right)e^{-i\lambda x},~~~
\widetilde{\psi}_{\lambda}(x;\lambda)\rightarrow i x \left(\begin{array}{c}
1 \\
0
\end{array}\right)e^{i\lambda x}.\\
\end{array}
\end{equation}
$\mathbf{Proof}$: A direct calculation gives rise to the conclusion.\\
$\mathbf{Lemma ~3.2.}$  When $\lambda \rightarrow \infty$, we have the following asymptotics
\begin{equation}
\begin{array}{l}
\phi(x;\lambda)\rightarrow \left(\begin{array}{c}
1 \\
0
\end{array}\right)e^{i\lambda x},~~~
\widetilde{\phi}(x;\lambda)\rightarrow \left(\begin{array}{c}
0 \\
1
\end{array}\right)e^{-i\lambda x},\\
\psi(x;\lambda)\rightarrow \left(\begin{array}{c}
0 \\
1
\end{array}\right)e^{-i\lambda x},~~~
\widetilde{\psi}(x;\lambda)\rightarrow \left(\begin{array}{c}
1 \\
0
\end{array}\right)e^{i\lambda x}.
\end{array}
\end{equation}
$\mathbf{Proof}$. From spectral problem \eqref{lax1}, we get
\begin{equation}
\left(\begin{array}{c}
\phi_{1}(x,\lambda) \\
\phi_{2}(x,\lambda)
\end{array}\right)_{x}=\left(\begin{array}{cc}
i \lambda+u & v \\
v & -i\lambda-u
\end{array}\right)\left(\begin{array}{c}
\phi_{1}(x,\lambda) \\
\phi_{2}(x,\lambda)
\end{array}\right),
\end{equation}
and also
\begin{align}
&\phi_{1,x}(x,\lambda)=(i\lambda+u)\phi_{1}(x,\lambda)+v\phi_{2}(x,\lambda),\label{bbb11}\\
&\phi_{2,x}(x,\lambda)=v\phi_{1}(x,\lambda)-(i\lambda+u)\phi_{2}(x,\lambda).\label{bb11}
\end{align}
By \eqref{bbb11}, we get
\begin{equation}\label{bb12}
\phi_{2}(x,\lambda)=\frac{\phi_{1,x}(x,\lambda)-(i\lambda+u)\phi_{1}(x,\lambda)}{v}.
\end{equation}
Substituting \eqref{bb12} into \eqref{bb11}, we obtain that
\begin{equation}
\left(\frac{\phi_{1, x}(x,\lambda)-(i\lambda+u)\phi_{1}(x,\lambda)}{v}\right)_{x}=v\phi_{1}(x,\lambda)-(i\lambda+u) \frac{\phi_{1,x}(x,\lambda)-(i\lambda+u)\phi_{1}(x,\lambda)}{v}.\label{bbb12} \\
\end{equation}
Rewriting equation \eqref{bbb12}, one also has
\begin{equation}\label{bb13}
\phi_{1, xx}(x,\lambda)-u_{x}\phi_{1}(x,\lambda)-\frac{\phi_{1,x}(x,\lambda)v_{x}}{v}+\frac{(i\lambda+u)
\phi_{1}(x,\lambda)v_{x}}{v}-v^{2}\phi_{1}(x,\lambda)+\lambda^{2}\phi_{1}(x,\lambda)-u^{2}\phi_{1}(x,\lambda)
-2i\lambda u \phi_{1}(x,\lambda)=0.
\end{equation}
Let\\
$~~~~~~~~~~~~~~~~~~~~~~~~~~~~~~~~~~~~~~~~~~~~~~~~~~~~~~~
\phi_{1}(x,\lambda)=e^{i\lambda x+\widehat{\phi}_{1}(x,\lambda)}$,\\
where\\
$~~~~~~~~~~~~~~~~~~~~~~~~~~~~~~~~~~~~~~~~~~~~~~~~~~~~~
\widehat{\phi}_{1}(x,\lambda)=\sum_{i=0}^{+\infty}\frac{a_{i}}{\lambda^{i}}$, $\lambda\rightarrow\infty$.\\
So we have
\begin{align}
&\phi_{1,x}(x,\lambda)=(i\lambda+\widehat{\phi}_{1,x}(x,\lambda))\phi_{1}(x,\lambda),\label{bb14}\\
&\phi_{1,xx}(x,\lambda)=\widehat{\phi}_{1,xx}(x,\lambda)\phi_{1}(x,\lambda)+
(i\lambda+\widehat{\phi}_{1,x}(x,\lambda))^{2}\phi_{1}(x,\lambda),\label{bb15}
\end{align}
substituting \eqref{bb14} and \eqref{bb15} into \eqref{bb13} yields
\begin{equation}\label{bb16}
\widehat{\phi}_{1, xx}(x,\lambda)+\widehat{\phi}_{1,x}^{2}(x,\lambda)+2i\lambda \widehat{\phi}_{1,x}(x,\lambda)-u_{x}-\frac{v_{x}}{v}\widehat{\phi}_{1,x}(x,\lambda)+
\frac{u}{v}v_{x}-v^2-u^2-2i\lambda u=0.
\end{equation}
By comparing the coefficients of $\lambda^{j}$ $(j=1,\cdots,n)$, we get
\begin{equation}
\begin{cases}
a_{0}=-\int_{x}^{+\infty}udx,\\
a_{1,x}=\frac{-iv^2}{2}.
\end{cases}
\end{equation}
Therefore, we get
\begin{equation}\label{bb20}
\phi_{1}(x,\lambda)=e^{i\lambda x-\int_{x}^{+\infty}(u-\frac{iv^2}{2\lambda})dx+o(\lambda^{-2})}.
\end{equation}
When $\lambda\rightarrow\infty$, we have
\begin{equation}\label{bb17}
\begin{aligned}
\phi_{2}(x,\lambda)&=\frac{\phi_{1,x}(x,\lambda)-(i\lambda+u)\phi_{1}(x,\lambda)}{v}\\
&\sim\frac{\left(i\lambda+u-\frac{iv^2}{2\lambda}\right)\phi_{1}(x,\lambda)-(i\lambda+u)\phi_{1}(x,\lambda)}{v}\\
&\sim\frac{-iv}{2\lambda}\phi_{1}(x,\lambda).
\end{aligned}
\end{equation}
Similarly, when $\lambda\rightarrow\infty$, we obtain that
\begin{equation}
\begin{cases}
\psi_{2}(x,\lambda)=e^{-i\lambda x-\int_{-\infty}^{x}(u+\frac{v^2}{2i\lambda})dx+o(\lambda^{-2})},\\
\psi_{1}(x,\lambda)=\frac{iv}{2\lambda}\psi_{2}(x,\lambda),
\end{cases}
\end{equation}
\begin{equation}
\begin{cases}
\widetilde{\phi}_{2}(x,\lambda)=e^{-i\lambda x+\int_{x}^{+\infty}(u+\frac{v^2}{2i\lambda})dx+o(\lambda^{-2})},\\
\widetilde{\phi}_{1}(x,\lambda)=\frac{iv}{2\lambda}\widetilde{\phi}_{2}(x,\lambda),
\end{cases}
\end{equation}
and
\begin{equation}\label{psy1}
\begin{cases}
\widetilde{\psi}_{1}(x,\lambda)=e^{i\lambda x+\int_{-\infty}^{x}(u-\frac{iv^2}{2\lambda})dx+o(\lambda^{-2})},\\
\widetilde{\psi}_{2}(x,\lambda)=\frac{-iv}{2\lambda}\widetilde{\psi}_{1}(x,\lambda)
.~~~~~~~~~~~~~~~~~~~~~~~~~~~ \Box
\end{cases}
\end{equation}
Define the Wronskian determinant of $\left( \phi(x,\lambda),\widetilde{\phi}(x,\lambda)\right)$ as
\begin{equation}
W(\phi(x,\lambda),\widetilde{\phi}(x,\lambda))=\phi_{1}(x,\lambda)\widetilde{\phi}_{2}(x,\lambda) -\widetilde{\phi}_{1}(x,\lambda)\phi_{2}(x,\lambda),
\end{equation}
and assume that
\begin{equation}
\begin{cases}
\phi(x,\lambda)=a(\lambda)\widetilde{\psi}(x,\lambda)+b(\lambda)\psi(x,\lambda),\\
\widetilde{\phi}(x,\lambda)=-\widetilde{a}(\lambda)\psi(x,\lambda)+\widetilde{b}(\lambda)\widetilde{\psi}(x,\lambda),\label{3c1}
\end{cases}
\end{equation}
 we have
\begin{equation}\label{b1}
\left(\begin{array}{cc}
\phi(x,\lambda) & \widetilde{\phi}(x,\lambda) \\
\end{array}\right)=
\left(\begin{array}{cc}
\widetilde{\psi}(x,\lambda) &  \psi(x,\lambda) \\
\end{array}\right)
\left(\begin{array}{cc}
 a(\lambda)  & \widetilde{b}(\lambda)  \\
 b(\lambda) & -\widetilde{a}(\lambda)
\end{array}\right).
\end{equation}
According to \eqref{lax1} and Lemma 3.1, we get
\begin{align}
&a(\lambda)=W(\phi(x;\lambda),\psi(x;\lambda)),~~\widetilde{a}(\lambda)=W(\widetilde{\phi}(x;\lambda),\widetilde{\psi}(x;\lambda)), \label{aa1} \\
&b(\lambda)=W(\widetilde{\psi}(x;\lambda),\phi(x;\lambda)),~~\widetilde{b}(\lambda)=W(\widetilde{\phi}(x;\lambda),\psi(x;\lambda)).
\end{align}
The solution of spectral problem \eqref{lax1} has the following asymptotics:
\begin{equation}\label{bb1}
\Psi(x,\lambda)=\left(\widetilde{\psi}(x,\lambda),\psi(x,\lambda) \right)\rightarrow e^{i\lambda x\sigma_{3}}, ~~x\rightarrow \infty.
\end{equation}
Taking the transformation
\begin{equation}
\mu(x,\lambda)=\Psi(x,\lambda)e^{-i\lambda x\sigma_{3}},
\end{equation}
we have
\begin{equation}
\mu(x,\lambda)\rightarrow I, ~~x\rightarrow \infty.
\end{equation}
From spectral problem \eqref{lax1}, we obtain
\begin{equation}\label{lax2}
\mu_{x}(x,\lambda)+i\lambda[\mu(x,\lambda),\sigma_{3}]=Q\mu(x,\lambda),
\end{equation}
where\\
$~~~~~~~~~~~~~~~~~~~~~~~~~~~~~~~~~~~~~~~~~~~~~~~~~~~~~~~~~~Q=\left(\begin{array}{cc}
u  & v  \\
v & -u
\end{array}\right)$.\\
Eq. \eqref{lax2} can be rewritten in differential form
\begin{equation}
d\left(e^{-i\lambda x\hat{\sigma}_{3}}\mu\right)=e^{-i\lambda x\hat{\sigma}_{3}\mu }[Qdx \mu].
\end{equation}
Therefore, we get
\begin{equation}
\mu_{-}=I+\int_{-\infty}^{x}e^{i\lambda (x-y)\hat{\sigma}_{3}}[Q\mu_{-} dy].
\end{equation}
$\mathbf{Lemma ~3.3.}$  $\mu_{-,1}(x,\lambda)$ and $\mu_{+,2}(x,\lambda)$ are analytical on $Im(\lambda)<0$, $\mu_{+,1}(x,\lambda)$ and $\mu_{-,2}(x,\lambda)$ are analytical on $Im(\lambda)>0$. So that $\phi(x,\lambda)$ and $\psi(x,\lambda)$ are analytical on $Im(\lambda)>0$, $\widetilde{\phi}(x,\lambda)$ and $\widetilde{\psi}(x,\lambda)$ are analytical on $Im(\lambda)<0$.\\
$\mathbf{Proof.}$ Let $\omega(x,\lambda)=\mu_{-,1}(x,\lambda)$, we have
\begin{equation}
\omega(x,\lambda)=\left(\begin{array}{c}
1  \\
0
\end{array}\right)
+\int_{-\infty}^{x}C_{-}(x,\lambda)\omega(y,\lambda)dy,
\end{equation}
where\\
$~~~~~~~~~~~~~~~~~~~~~~~~~~~~~~~~~~~~~~~~~~~~~~~~~~C_{-}(x,\lambda)=diag(1,e^{-2i\lambda(x-y)})Q(y,\lambda)$.\\
We constructe a Neumann series
\begin{equation}\label{bb6}
\omega(x,\lambda)=\sum_{n=1}^{\infty}\omega^{(n)}(x,\lambda),
\end{equation}
and define the norm of the vector $L^{1}$ as $\|\omega \|=|\omega_{1}|+|\omega_{2}|$. A direct calculation yields the following estimation
\begin{equation}\label{bb2}
\|C_{-}\|=\|diag(1,e^{-2i\lambda(x-y)})\| \cdot \|Q\|\leq \left(1+e^{2Im\lambda (x-y)}\right)\cdot(2|u|+2|v|)\leq 4(|u|+|v|).
\end{equation}
So we have
\begin{align}
&\omega^{0}=\left(\begin{array}{c}
1  \\
0
\end{array}\right),\\
&\omega^{n+1}(x,\lambda)=\int_{-\infty}^{x}C_{-}(y,z)\omega^{n}(y,\lambda)dy,\label{bb5}
\end{align}
which indicates
\begin{equation}\label{bb3}
\|\omega^{(n+1)}(x,\lambda)\|\leq \int_{-\infty}^{x}\|C_{-}\|\|\omega^{(n)}(y,\lambda)\|dy.
\end{equation}
Substituting \eqref{bb2} into \eqref{bb3}, it can be obtained that
\begin{equation}\label{bb4}
\|\omega^{n+1}(x,\lambda)\|\leq \int_{-\infty}^{x}4(|u|+|v|)\|\omega^{(n)}(y,\lambda)\|dy.
\end{equation}
We denote\\
$~~~~~~~~~~~~~~~~~~~~~~~~~~~~~~~~~~~~~~~~~~~~~~~~~~~~\rho(x)=\int_{-\infty}^{x}4(|u|+|v|)dy$,\\
the above integral exists when $u(x)\in L^{1}(R)$, $v(x)\in L^{1}(R)$, for any fixed $x \in R$, we have
\begin{equation}
\rho_{x}(x)=4(|u|+|v|),~~\|\omega^{(1)}\|\leq \rho(x).
\end{equation}
By mathematical induction and assuming that $\|\omega^{(n)}(x,\lambda)\|\leq \frac{\rho^{n}(x)}{n!}$, we obtain that
\begin{equation}
\|\omega^{(n+1)}(x,\lambda)\|\leq \int_{-\infty}^{x}\rho_{x}(x)\frac{\rho^{n}(x)}{n!}=\frac{\rho^{n+1}(x)}{(n+1)!}.
\end{equation}
 Therefore, $\omega^{(n)}(x,\lambda)$ defined by \eqref{bb5} exists and is bounded, and additionally by the analyticity of $\omega^{(0)}$ and $C_{-}$ with respect to $\lambda$, recursively $\omega^{(n)}(x,\lambda),n\geq 1$ are analytical.

For $x\leq a \in R$, we get\\
$~~~~~~~~~~~~~~~~~~~~~~~~~~~~~~~~~~~~~~~~~~~~~~~~~~~~~~~~~\rho(x)\leq \int_{-\infty}^{a}4(|u|+|v|)dy=\sigma$,\\
where $a$ is a bounded real constant and $\sigma$ is a constant independent of $x$. Then we have
\begin{equation}
\|\omega^{(n)}(x,\lambda)\|\leq \frac{\sigma^{n}}{n!}
\end{equation}
and\\
$~~~~~~~~~~~~~~~~~~~~~~~~~~~~~~~~~~~~~~~~~~~~~~~~~~~\|\sum_{n=0}^{+\infty}\omega^{(n)}(x,\lambda)\|\leq \sum_{n=0}^{+\infty}\frac{\sigma^{n}}{n!}=e^{a}$.\\
Therefore, the series \eqref{bb6} converges absolutely for $x\in (-\infty,a)$, and
\begin{equation}\label{bb7}
\omega(x,\lambda)=\sum_{n=0}^{+\infty}\omega^{(n)}(x,\lambda),
\end{equation}
so $\omega(x,\lambda)$ is analytical on $Im(\lambda)<0$.\\
From \eqref{bb5} and \eqref{bb7}, we obtain
\begin{equation}\label{bb8}
\begin{aligned}
&~~~\omega(x,\lambda)\\
&=\omega^{(0)}+\sum_{n=1}^{+\infty}\omega^{(n)}(x,\lambda)\\
&=\omega^{(0)}+\int_{-\infty}^{x}C_{-}\sum_{n=0}^{+\infty}\omega^{(n)}(y,\lambda)dy\\
&=\omega^{(0)}+\int_{-\infty}^{x}C_{-}\omega(y,z)dy,
\end{aligned}
\end{equation}
this equation shows that $\omega(x,\lambda)$ defined by \eqref{bb7} is a solution of equation \eqref{bb5}.

To prove the uniqueness of the solution $\omega(x,\lambda)$, we assume that $\widetilde{\omega}(x,\lambda)$ is another solution of equation \eqref{bb5}. Let $h(x,\lambda)=\widetilde{\omega}(x,\lambda)-\omega(x,\lambda)$, we obtain\\
$~~~~~~~~~~~~~~~~~~~~~~~~~~~~~~~~~~~~~~~~~~~~~~~\|h(x,\lambda)\|\leq 4\int_{-\infty}^{x}(|u|+|v|)\|h(y,\lambda)\|dy$.\\
Accordong to Bellmann's inequality, we have $h(x,\lambda)=0$, so $\omega(x,\lambda)$ defined by \eqref{bb7} is the unique solution of \eqref{bb5}.
~~~~~~~~~~~~~~~~~~~~~~~~~~~~~~~~~~~~~~~~~~~~~~~~~~~~~~~~~~~~~~~~~~~~
~~~~~~~~~~~~~~~~~~~~~~~~~~~~~~~~~~~~~~~~~~~~~~~~~~~~~~~~~~$\Box$\\
$\mathbf{Remark~ 3.1.}$ Similiarly, we can obtain $\mu_{+,2}(x,\lambda)$ is analytical on $Im(\lambda)<0$, $\mu_{+,1}(x,\lambda)$ and $\mu_{-,2}(x,\lambda)$ are analytical on $Im(\lambda)>0$.

\subsection{GLM equation}
Supposing that $a(\lambda)$ has $N$ simple zeros $\lambda_{1}, \lambda_{2}, \cdots \lambda_{N}$ on
$Im(\lambda)>0$, and $\widetilde{a}(\lambda)$ has $N$ simple zeros $\widetilde{\lambda}_{1}, \widetilde{\lambda}_{2}, \cdots \widetilde{\lambda}_{N}$ on $Im(\lambda)<0$,
it follows that
\begin{equation}\label{bbb2}
a(\lambda_{j})=0, (j=1,2,\cdots ,N)
\end{equation}
and
\begin{equation}\label{bbbb2}
\widetilde{a}(\widetilde{\lambda}_{j})=0, (j=1,2,\cdots ,N).
\end{equation}
From Lemma 3.2, \eqref{bbb2} and \eqref{bbbb2}, we get
\begin{equation}\label{bbb1}
\phi(x,\lambda_{j})=b_{j}\psi(x,\lambda_{j}),
\end{equation}
and
\begin{equation}\label{bbbb1}
\widetilde{\phi}(x,\widetilde{\lambda}_{j})=\widetilde{b}_{j}\widetilde{\psi}(x,\widetilde{\lambda}_{j}),
\end{equation}
where $b_{j}$ and $\widetilde{b}_{j}$ are constants.
In this paper, we consider the case of simple zeros, that is,
\begin{equation}
a^{'}(\lambda_{j})=\frac{d}{d\lambda}a(\lambda)|_{\lambda=\lambda_{j}}\neq 0.
\end{equation}
So we have
\begin{equation}
a(\lambda)=\prod_{j=1}^{N}\frac{\lambda-\lambda_{j}}{\lambda-\overline{\lambda}_{j}}\exp\left\{\frac{1}{2\pi i}\int_{-\infty}^{+\infty}\frac{\ln |a(\lambda^{'})|^{2}}{\lambda^{'}-\lambda}d\lambda^{'}\right\}.
\end{equation}
According to Lemma 3.3, we define
\begin{equation}
\Theta(x,\lambda)=\begin{cases}
a^{-1}(\lambda)\phi(x,\lambda),~~Im\lambda >0,\\
\widetilde{\psi}(x,\lambda),~~~~~~~~~~~Im\lambda <0.
\end{cases}
\end{equation}
$a^{-1}(\lambda)\phi(x,\lambda)$ is analytical on $Im(\lambda)>0$ except for $\lambda_{j}$, $\widetilde{\psi}(x,\lambda)$ is analytical on $Im(\lambda)<0$ except for $\widetilde{\lambda}_{j}$, $\Theta(x,\lambda)$ has jumps at the real axis that
\begin{equation}
a^{-1}(\lambda)\phi(x,\lambda)-\widetilde{\psi}(x,\lambda)=r(\lambda)\psi(x,\lambda),
\end{equation}
where\\
$~~~~~~~~~~~~~~~~~~~~~~~~~~~~~~~~~~~~~~~~~~~~~~~~~~~~~~~~~~~~~~~~r(\lambda)=\frac{b(\lambda)}{a(\lambda)}$.\\
From \eqref{aa1}, we get
\begin{equation}\label{bb17}
\begin{aligned}
a(\lambda)&=\phi_{1}(x,\lambda)\psi_{2}(x,\lambda)-\psi_{1}(x,\lambda)\phi_{2}(x,\lambda)\\
&=\psi_{2}(x,\lambda)\phi_{1}(x,\lambda)-\frac{iv}{2\lambda}\psi_{2}(x,\lambda)\cdot
\left(\frac{-iv}{2\lambda}\phi_{2}(x,\lambda)\right)\\
&=\psi_{2}(x,\lambda)\phi_{1}(x,\lambda)-\frac{v^2}{4\lambda^{2}}\psi_{2}(x,\lambda)\phi_{1}(x,\lambda)\\
&=e^{-\int_{-\infty}^{+\infty}\left(u-\frac{iv^2}{2\lambda}\right)dx}+o\left(\frac{1}{\lambda^{2}}\right).
\end{aligned}
\end{equation}
So we have\\
$~~~~~~~~~~~~~~~~~~~~~~~~~~~~~~~~~~~~~~~~~~~~~~~~~\frac{\phi_{1}(x,\lambda)}{a(\lambda)}=e^{i\lambda x}\cdot e^{\int_{-\infty}^{x}(u-\frac{iv^2}{2\lambda})dx}$,~~
$\frac{\phi_{2}(x,\lambda)}{a(\lambda)}=\frac{-iv}{2\lambda}\frac{\phi_{1}(x,\lambda)}{a(\lambda)}$\\
and\\
$~~~~~~~~~~~~~~~~~~~~~~~~~~~~~~~~~~~~~~~~~~~~~~\frac{\phi_{1}(x,\lambda)}{a(\lambda)}e^{-i\lambda x}=e^{\int_{-\infty}^{x}udx}+o(\frac{1}{\lambda})$,~~
$\frac{\phi_{2}(x,\lambda)}{a(\lambda)}e^{-i\lambda x}=o(\frac{1}{\lambda})$.\\
In addition, \\
$~~~~~~~~~~~~~~~~~~~~~~~~~~~~~~~~~~~~~~~~~~~~~~\widetilde{\psi}_{1}(x,\lambda)e^{-i\lambda x}=e^{\int_{-\infty}^{x}udx}+o(\frac{1}{\lambda})$,~~
$\widetilde{\psi}_{2}(x,\lambda)e^{-i\lambda x}=o(\frac{1}{\lambda})$.\\
Therefore, there are
\begin{equation}\label{wt2}
\Theta(x,\lambda) e^{-i\lambda x-\int_{-\infty}^{x}udx}-\left(\begin{array}{c}
1 \\
0
\end{array}\right)=o\left(\frac{1}{\lambda}\right),
\end{equation}
applying Cauchy's formula, we get
\begin{equation}\label{the1}
\Theta(x,\lambda)e^{-i\lambda x-\int_{-\infty}^{x}udx}-\left(\begin{array}{c}
1 \\
0
\end{array}\right)=\frac{1}{2\pi i}\oint \frac{\Theta(x,\lambda^{'})e^{-i\lambda^{'} x-\int_{-\infty}^{x}udx}-\left(\begin{array}{c}
1 \\
0
\end{array}\right)}{\lambda^{'}-\lambda}d\lambda^{'}.
\end{equation}
where the integration path is shown in Fig. 1 and consists of two counterclockwise semicircles with radius tending to infinity and $N$ small clockwise circles $r_j, j=1,2,...N.$ containing the poles $\lambda_j, j=1,2,...N.$ of $\Theta(x,\lambda)$.

\begin{figure}
  \centering
  \includegraphics[width=8cm]{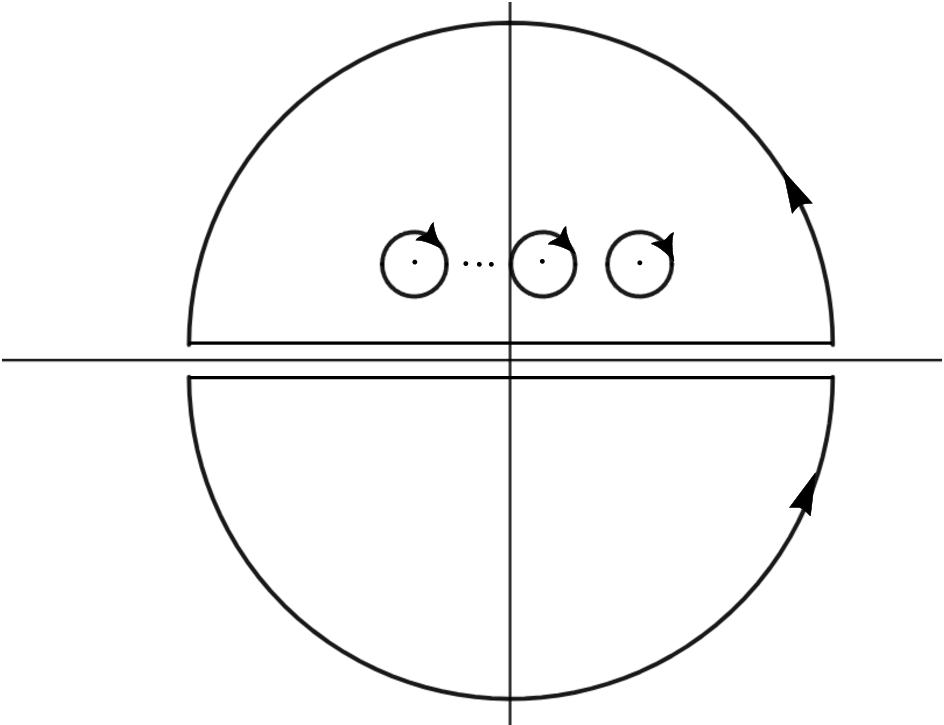}
  \caption{The integration path in \eqref{the1}.}
\end{figure}

We note that the integral over the two arc of the great semicircle tends to $0$ since \eqref{wt2}. The right end of the above equation reduces to the sum of the residues
\begin{equation}
R(x,\lambda)=\frac{1}{2\pi i}\oint_{r_{j}}\frac{\Theta(x,\lambda^{'})e^{-i\lambda^{'} x-\int_{-\infty}^{x}udx}-\left(\begin{array}{c}
1 \\
0
\end{array}\right)}{\lambda^{'}-\lambda}d\lambda^{'}
\end{equation}
 at $\lambda_{j}$, and the continuous spectral component $J(x,\lambda)$
\begin{equation}
J(x,\lambda)=\frac{1}{2\pi i}\int_{-\infty}^{+\infty}\frac{r(\lambda^{'})\psi(x,\lambda^{'})e^{-i\lambda^{'}x-\int_{-\infty}^{x}udx}}{\lambda^{'}-\lambda}d\lambda^{'}.
\end{equation}
Applying \eqref{bbb1}, one obtains
\begin{equation}
R(x,\lambda)=\sum_{j=1}^{N}\frac{i}{\lambda-\lambda_{j}}c_{j}\psi(x,\lambda_{j})e^{-i\lambda_{j}x-\int_{-\infty}^{x}udx},
\end{equation}
where $c_{j}=\frac{b_{j}}{a^{'}(\lambda_{j})}$.\\
For the reflectionless case, when $Im(\lambda)<0$, have
\begin{equation}\label{glm1}
\left(\begin{array}{c}
\widetilde{\psi}_{1}(x,\lambda) \\
\widetilde{\psi}_{2}(x,\lambda)
\end{array}\right)\cdot e^{-i\lambda x-\int_{-\infty}^{x}udx}
-\left(\begin{array}{c}
1 \\
0
\end{array}\right)=\sum_{j=1}^{N}\frac{i}{\lambda-\lambda_{j}} c_{j}\left(\begin{array}{c}
\psi_{1}(x,\lambda_{j}) \\
\psi_{2}(x,\lambda_{j})
\end{array}\right)e^{-i\lambda_{j}x-\int_{-\infty}^{x}u dx}.
\end{equation}
In addition, from \eqref{3c1}, we have\\
$~~~~~~~~~~~~~~~~~~~~~~~~~~~~~~~~~~~~~~~~~~~~~~~~~~\psi(\lambda)-(-\widetilde{a}(\lambda)^{-1}\widetilde{\phi}(\lambda))=\widetilde{r}(\lambda)\widetilde{\psi}(\lambda)$,\\
where $\widetilde{r}(\lambda)=\frac{\widetilde{b}(\lambda)}{\widetilde{a}(\lambda)}$.

Define
\begin{equation}
\widetilde{\Theta}(x,\lambda)=\begin{cases}
\phi(x,\lambda),~~~~~~~~~~~~~~Im\lambda >0,\\
-\widetilde{a}(\lambda)^{-1}\widetilde{\psi}(x,\lambda),~~~Im\lambda <0.
\end{cases}
\end{equation}
From \eqref{aa1}, we know that
\begin{equation}\label{bb17}
\begin{aligned}
\widetilde{a}(\lambda)&=\widetilde{\phi}_{1}(x,\lambda)\widetilde{\psi}_{2}(x,\lambda)
-\widetilde{\psi}_{1}(x,\lambda)\widetilde{\phi}_{2}(x,\lambda)\\
&=\left(\frac{v^2}{4\lambda^2}-1\right)\widetilde{\psi}_{1}(x,\lambda)\widetilde{\phi}_{2}(x,\lambda)\\
&=-e^{\int_{-\infty}^{+\infty}\left(u-\frac{iv^2}{2\lambda}\right)dx}+o\left(\frac{1}{\lambda^2}\right).
\end{aligned}
\end{equation}
Furthermore, we have
\begin{equation}
\begin{cases}
\psi_{1}(x,\lambda)e^{i\lambda x+\int_{-\infty}^{x}u dx}=o\left(\frac{1}{\lambda}\right),\\
\psi_{2}(x,\lambda)e^{i\lambda x+\int_{-\infty}^{x}u dx}-1=o\left(\frac{1}{\lambda}\right),\\
\frac{\widetilde{\phi}_{2}(x,\lambda)}{-\widetilde{a}(\lambda)}e^{i\lambda x+\int_{-\infty}^{x}u dx}-1=o\left(\frac{1}{\lambda}\right),\\
\frac{\widetilde{\phi}_{1}(x,\lambda)}{-\widetilde{a}(\lambda)}e^{i\lambda x+\int_{-\infty}^{x}u dx}=o\left(\frac{1}{\lambda}\right),
\end{cases}
\end{equation}
which in turn yields
\begin{equation}\label{wt1}
\widetilde{\Theta}(x,\lambda)e^{i\lambda x+\int_{-\infty}^{x}u dx}-\left(\begin{array}{c}
0 \\
1
\end{array}\right)=o\left(\frac{1}{\lambda}\right).
\end{equation}
According to Cauchy's formula, we get
\begin{equation}\label{the2}
\widetilde{\Theta}(x,\lambda)\cdot e^{i\lambda x+\int_{-\infty}^{x}u dx}-\left(\begin{array}{c}
0 \\
1
\end{array}\right)=\frac{1}{2\pi i}\oint \frac{\widetilde{\Theta}(x,\lambda^{'})e^{i\lambda^{'} x+\int_{-\infty}^{x}u dx}-\left(\begin{array}{c}
0 \\
1
\end{array}\right)}{\lambda^{'}-\lambda}d\lambda^{'}.
\end{equation}
where the integration path is shown in Fig. 2 and consists of two counterclockwise semicircles with radius tending to infinity and $N$ small clockwise circles $\widetilde{r}_j, j=1,2,...N.$ containing the poles $\widetilde{\lambda}_j, j=1,2,...N.$ of $\widetilde{\Theta}(x,\lambda)$.

\begin{figure}
  \centering
  \includegraphics[width=8cm]{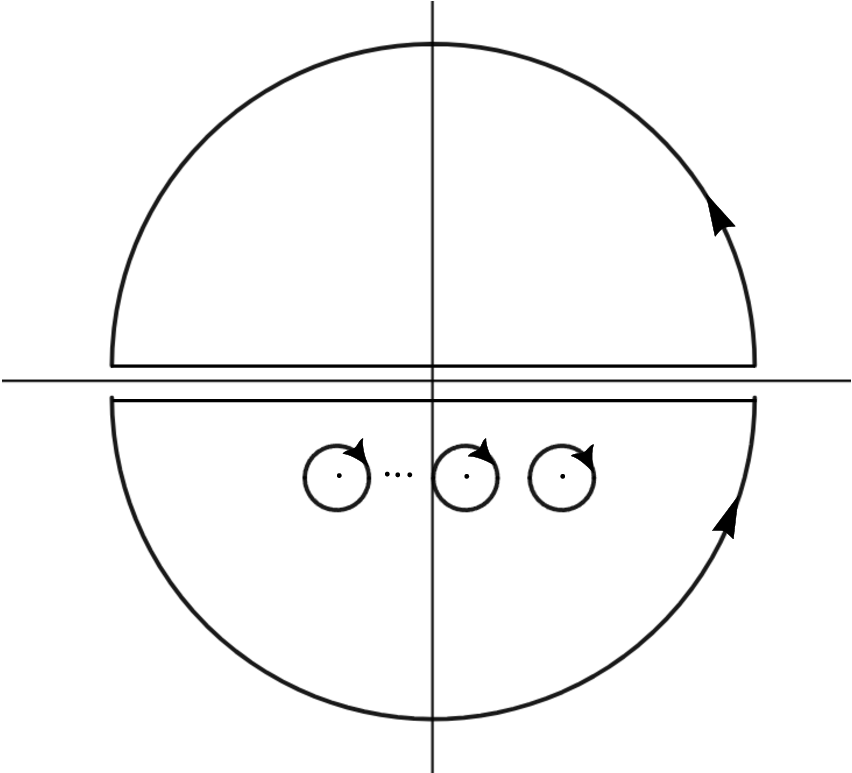}
  \caption{The integration path in \eqref{the2}.}
\end{figure}

We note that the integral over the two arc of the great semicircle tends to $0$ since \eqref{wt1}. The right end of the above equation reduces to the sum of the residues
\begin{equation}
\widetilde{R}(x,\lambda)=\frac{1}{2\pi i}\oint_{\widetilde{r}_{j}}\frac{\widetilde{\Theta}(x,\lambda^{'})e^{i\lambda^{'} x+\int_{-\infty}^{x}udx}-\left(\begin{array}{c}
1 \\
0
\end{array}\right)}{\lambda^{'}-\lambda}d\lambda^{'}
\end{equation}
 at $\lambda_{j}$, and the continuous spectral component $J(x,\lambda)$
\begin{equation}
\widetilde{J}(x,\lambda)=\frac{1}{2\pi i}\int_{-\infty}^{+\infty}\frac{\widetilde{r}(\lambda^{'})\widetilde{\psi}(x,\lambda^{'})e^{i\lambda^{'}x+\int_{-\infty}^{x}udx}}{\lambda^{'}-\lambda}d\lambda^{'}.
\end{equation}
Through \eqref{bbbb1}, we obtain
\begin{equation}
\widetilde{R}(x,\lambda)=\sum_{j=1}^{N}\frac{i}{\lambda-\widetilde{\lambda}_{j}}\tilde{c}_{j}\widetilde{\psi}(x,\widetilde{\lambda}_{j})e^{i\widetilde{\lambda}_{j}x-\int_{-\infty}^{x}udx},
\end{equation}
where $\widetilde{c}_{j}=-\frac{\widetilde{b}_{j}}{\widetilde{a}^{'}(\widetilde{\lambda}_{j})}$.\\
For the reflectionless case, when $Im(\lambda)>0$,
\begin{equation}\label{glm2}
\left(\begin{array}{c}
\psi_{1}(x,\lambda) \\
\psi_{2}(x,\lambda)
\end{array}\right)e^{i\lambda x+\int_{-\infty}^{x}u dx}-\left(\begin{array}{c}
0 \\
1
\end{array}\right)=
\sum_{j=1}^{N}\frac{i}{\lambda-\widetilde{\lambda}_{j}}\tilde{c}_{j}\left(\begin{array}{c}
\widetilde{\psi}_{1}(x,\widetilde{\lambda}_{j}) \\
\widetilde{\psi}_{2}(x,\widetilde{\lambda}_{j})
\end{array}\right)e^{i\widetilde{\lambda}_{j}x+\int_{-\infty}^{x}u dx}.
\end{equation}

\subsection{The time dependence of the scattering data}
In this subsection, we determine the time evolution of the non-isospectral scattering data.

From \eqref{lax1} and \eqref{aa1} we know that $a(\lambda)=\phi_{1}(x,\lambda)\psi_{2}(x,\lambda)-\psi_{1}(x,\lambda)\phi_{2}(x,\lambda)$, also have
\begin{equation}\label{bb8}
\begin{aligned}
a_{\lambda}(\lambda_{j})&=\left(\psi_{2,\lambda}(x,\lambda_{j})\phi_{1}(x,\lambda_{j})
+\psi_{2}(x,\lambda_{j})\phi_{1,\lambda}(x,\lambda_{j})\right)-\left(\psi_{1,\lambda}(x,\lambda_{j})
\phi_{2}(x,\lambda_{j})+\psi_{1}(x,\lambda_{j})\phi_{2,\lambda}(x,\lambda_{j})\right)\\
&=\int_{-\infty}^{x}\frac{\partial}{\partial x}\left(\psi_{1,\lambda}(x,\lambda_{j})\phi_{2}(x,\lambda_{j})+
\psi_{1}(x,\lambda_{j})\phi_{2,\lambda}(x,\lambda_{j})\right)dx\\
&~~~-\int_{x}^{+\infty}\frac{\partial}{\partial x}\left(\psi_{2,\lambda}(x,\lambda_{j})\phi_{1}(x,\lambda_{j})+\psi_{2}(x,\lambda_{j})
\phi_{1,\lambda}(x,\lambda_{j})\right)dx\\
&=\int_{-\infty}^{+\infty}2ib_{j}\psi_{1}(x,\lambda_{j})\psi_{2}(x,\lambda_{j}).
\end{aligned}
\end{equation}
Using $\sqrt{ic_{j}}\psi_{1}(x,\lambda_{j})$ instead of $\psi_{1}(x,\lambda_{j})$, $\sqrt{ic_{j}}\psi_{2}(x,\lambda_{j})$ instead of $\psi_{2}(x,\lambda_{j})$, we get
\begin{equation}\label{define3.1}
\int_{-\infty}^{+\infty}2\psi_{1}(x,\lambda_{j})\psi_{2}(x,\lambda_{j})=1.
\end{equation}
$\mathbf{Lemma ~3.4.}$ Suppose that $\psi(x,t,\lambda)$ is a solution of spectral problem \eqref{lax1}, $M$ and $N$ satisfy the zero curvature equation \eqref{zero1}, then
\begin{equation}\label{lem3.1}
P(x,t,\lambda)=\psi_{t}(x,t,\lambda)-N\psi(x,t,\lambda)
\end{equation}
is a solution of spectral problem \eqref{lax1} as well.\\
$\mathbf{Proof}$. By direct calculation, we get\\
$~~~~~~~~~~~~~~~~~~~~~~~~~~~~~~~~~~~~~~~~~P_x(x,t,\lambda)=\psi_{tx}(x,t,\lambda)-N_{x}\psi(x,t,\lambda)-N\psi_{x}(x,t,\lambda)$\\
$~~~~~~~~~~~~~~~~~~~~~~~~~~~~~~~~~~~~~~~~~~~~~~~~~~~~~~=(M\psi (x,t,\lambda))_{t}-N_{x}\psi (x,t,\lambda)-N\psi_{x}(x,t,\lambda)$\\
$~~~~~~~~~~~~~~~~~~~~~~~~~~~~~~~~~~~~~~~~~~~~~~~~~~~~~~=M_{t}\psi(x,t,\lambda)+M\psi_{t}(x,t,\lambda)-N_{x}\psi(x,t,\lambda)
-N\psi_{x}(x,t,\lambda)$\\
$~~~~~~~~~~~~~~~~~~~~~~~~~~~~~~~~~~~~~~~~~~~~~~~~~~~~~~=MP(x,t,\lambda)$.\\
Therefore, $P(x,t,\lambda)$ is a solution of spectral problem \eqref{lax1}.  ~~~~~~~~~~~~~~~~~~~~~~~~~~~~~~~~~~~~~~~~~~$\Box$   \\
$\mathbf{Definition ~3.1.}$ For brevity, we define the vector operations $\cdot$ and $\langle, \rangle$ as\\
$~~~~~~~~~~~~~~~~~~~~~~~~~~~~~~~~~~~~~~\left(\begin{array}{c}
z_{1} \\
z_{2}
\end{array}\right)\cdot\left(\begin{array}{c}
y_{1} \\
y_{2}
\end{array}\right)=z_{1}y_{1}+z_{2}y_{2}$,\\
$~~~~~~~~~~~~~~~~~~~~~~~~~~~~~~~~~~~~~~\left\langle \left(\begin{array}{c}
z_{1} \\
z_{2}
\end{array}\right),\left(\begin{array}{c}
y_{1} \\
y_{2}
\end{array}\right) \right\rangle=\int_{-\infty}^{+\infty}\left(z_{1}y_{1}+z_{2}y_{2}\right)dx$.\\
$\mathbf{Theorem~ 3.1.}$ The discrete scattering data ${\lambda_{j}(t), c_{j}(t)}$ for $\psi(x,t,\lambda)$ and $\phi(x,t,\lambda)$ have the following time dependence. For $n=0$,
\begin{equation}
\begin{cases}
\lambda_{j}(t)=e^{t+\ln \lambda_{j}},\\
c_j(t)=e^{t+\ln c_j(t)},
\end{cases}
\end{equation}
For $n\geq 1$,
\begin{equation}
\begin{cases}
\lambda_{j}(t)=\left(\lambda_{j}(0)^{-n}-nt\right)^{-\frac{1}{n}},\\
c_j(t)=\frac{c_{j}(0)}{\lambda_{j}(0)^{n+1}}\left(\lambda_{j}(0)^{-n}-nt\right)^{-\frac{n+1}{n}},
\end{cases}
\end{equation}
where $\lambda_j(0)$ and $c_j(0)$ are the scattering data of spectral problem \eqref{lax1} in terms of $u(0, x)$ and $v(0, x)$.\\
$\mathbf{Proof}$. Taking $\lambda=\lambda_j$ (where $Im\lambda_j>0)$ and replacing the normalization eigenfunction $\sqrt{ic_{j}} \psi(x,\lambda_j)$ of spectral problem \eqref{lax1} by $\psi(x,\lambda_j)$, and based on Lemma 3.4, we know\\
$~~~~~~~~~~~~~~~~~~~~~~~~~~~~~~~~~~~~~~~~~~~~~~~~~~~~P(x,\lambda_j)=\psi_t(x,\lambda_j)-N\psi(x,\lambda_j)$\\
is still a solution of spectral problem \eqref{lax1}.
There exist two constants $\beta$ and $\theta$, such that
\begin{equation}\label{th3.1}
\psi_{t}(x,\lambda_j)-N\psi(x,\lambda_j)=\beta \psi(x,\lambda_j)+\theta \widetilde{\psi}(x,\lambda_j).
\end{equation}
We can find $\theta=0$ because $\psi_1(x,\lambda_j)$ tends to zero while $\widetilde{\psi}_1(x,\lambda_j)$ tends to infinite when
$x\rightarrow +\infty$. So \eqref{th3.1} reads
\begin{equation}\label{th3.11}
\psi_{t}(x,\lambda_j)-N\psi(x,\lambda_j)=\beta \psi(x,\lambda_j).
\end{equation}
To derive the value of $\beta$, multiplying \eqref{th3.11} by $\left(\psi_2(x,\lambda_{j}),\psi_1(x,\lambda_{j})\right)^T$ ($T$ is the transpose of the matrix), we get
\begin{equation}\label{th3.12}
\begin{aligned}
&\int_{-\infty}^{+\infty} [2(\psi_1(x,\lambda_{j})\psi_2(x,\lambda_{j}))_{t}-A\psi_{1}(x,\lambda_{j})\psi_{2}(x,\lambda_{j})
-(B+C)\psi_{2}^{2}(x,\lambda_{j})\\
&-(B-C)\psi_{1}^{2}(x,\lambda_{j})+A\psi_{2}(x,\lambda_{j})\psi_{1}(x,\lambda_{j})
]dx=\beta.
\end{aligned}
\end{equation}
Through \eqref{define3.1}, we obtain
\begin{equation}\label{3a2}
\int_{-\infty}^{+\infty}\left(-(B+C)\psi_{2}^{2}(x,\lambda_{j})-(B-C)\psi_{1}^{2}(x,\lambda_{j})\right)dx=\beta.
\end{equation}
Calculating \eqref{3a2} yield
\begin{equation}
\begin{aligned}
\beta &=-\int_{-\infty}^{+\infty}\left(\begin{array}{l}
B+C \\
B-C
\end{array}\right)\left(\begin{array}{l}
\psi_{2}^{2}(x,\lambda_{j}) \\
\psi_{1}^{2}(x,\lambda_{j})
\end{array}\right) d x \\
&=-\int_{-\infty}^{+\infty} \sum_{i=0}^{n}\left(\begin{array}{c}
b_{i+1}+c_{i} \\
b_{i+1}-c_{i}
\end{array}\right)\left(\begin{array}{l}
\psi_{2}^{2}(x,\lambda_{j}) \\
\psi_{1}^{2}(x,\lambda_{j})
\end{array}\right)dx \cdot \lambda_{j}^{n-i} \\
&=-(n+1)\int_{-\infty}^{+\infty} b_{1}\left(\psi_{2}^{2}(x,\lambda_{j})+\psi_{1}^{2}(x,\lambda_{j})\right) \lambda_{j}^{n}dx \\
&=-(n+1) \int_{-\infty}^{+\infty}vx\left(\psi_{1}^{2}(x,\lambda_{j})+\psi_{2}^{2}(x,\lambda_{j})\right)
 \lambda_{j}^{n}dx \\
&=(n+1)\int_{-\infty}^{+\infty}\psi_{1}(x,\lambda_{j})\psi_{2}(x,\lambda_{j}) dx\lambda_{j}^{n}\\
&=\frac{\lambda_{j}^{n}}{2} \cdot(n+1).
\end{aligned}
\end{equation}
From \eqref{asym2} and \eqref{th3.11}, we get
\begin{equation}
\begin{aligned}
&\psi_{t}(x,\lambda_{j})-N\psi(x,\lambda_{j})\\
=&-ix\lambda_{j,t}\left(\begin{array}{cc}
0 \\
1
\end{array}\right)e^{-i\lambda_{j} x}\sqrt{ic_{j}}+(\sqrt{ic_{j}})_{t}\left(\begin{array}{l}
0 \\
1
\end{array}\right)e^{-i\lambda_{j} x}\\
&-\left(\begin{array}{cc}
ix\lambda_{j}^{n+1}  &  0 \\
0  &  -ix\lambda_{j}^{n+1}
\end{array}\right)
\left(\begin{array}{l}
0 \\
1
\end{array}\right)e^{-i\lambda_{j} x}\sqrt{ic_{j}}\\
=&\frac{n+1}{2}\lambda_{j}^{n}\left(\begin{array}{l}
0 \\
1
\end{array}\right)e^{-i\lambda_{j} x}\sqrt{ic_{j}}.
\end{aligned}
\end{equation}
as $x\rightarrow-\infty$.
It can be deduced that\\
$~~~~~~~~~~~~~~~~~~~~-ix\lambda_{j,t}\left(\begin{array}{l}
0 \\
1
\end{array}\right)e^{-i\lambda_{j}x}\sqrt{ic_{j}}=-\left(\begin{array}{cc}
ix\lambda_{j}^{n+1}  &  0 \\
0  &  -ix\lambda_{j}^{n+1}
\end{array}\right)\left(\begin{array}{l}
0 \\
1
\end{array}\right)e^{-i\lambda_{j} x}\sqrt{ic_{j}}$\\
and\\
$~~~~~~~~~~~~~~~~~~~~~~~~~~~~~~~~~~~~~~~~~~~~~~~~~~~~~~~~~~~(\sqrt{ic_{j}})_{t}=\frac{n+1}{2}\lambda^{n}\sqrt{ic_{j}}$.\\
Thus we have \\
$~~~~~~~~~~~~~~~~~~~~~~~~~~~~~~~~~~~~~~~~~~~~~~~~~~~~~~~~~~~~~~c_{j,t}=(n+1)\lambda_{j}^{n}c_{j}$. \\
Since $\lambda_t=\lambda^{n+1}$, we can prove the theorem directly.
~~~~~~~~~~~~~~~~~~~~~~~~~~~~~~~~~~~~~~~~~~~~~~~~~~~~~~~~~~~~~~~~~~~~~~~~ $\Box$

On the other hand, from \eqref{aa1}, we derive that\\
$~~~~~~~~~~~~~~~~~~~~~~~~~~~~~~~~~~~~~~~~~~~~~~~~~~~~~~~~~~~~~
\widetilde{a}(\lambda)=\widetilde{\phi}_{1}(x,\lambda)\widetilde{\psi}_{2}(x,\lambda)
-\widetilde{\psi}_{1}(x,\lambda)\widetilde{\phi}_{2}(x,\lambda)$.\\
In this case,
\begin{equation}
\begin{aligned}
\widetilde{a}_{\lambda}(\lambda)&=\left(\widetilde{\phi}_{1,\lambda}(x,\lambda)\widetilde{\psi}_{2}(x,\lambda)
+\widetilde{\phi}_{1}(x,\lambda)\widetilde{\psi}_{2,\lambda}(x,\lambda)\right)
-\left(\widetilde{\psi}_{1,\lambda}(x,\lambda)\widetilde{\phi}_{2}(x,\lambda)+\widetilde{\psi}_{1}(x,\lambda)
\widetilde{\phi}_{2,\lambda}(x,\lambda)\right)\\
&=\int_{-\infty}^{x}\frac{\partial}{\partial x}\left(\widetilde{\phi}_{1,\lambda}(x,\lambda)\widetilde{\psi}_{2}(x,\lambda)-\widetilde{\psi}_{1}(x,\lambda)
\widetilde{\phi}_{2,\lambda}(x,\lambda)\right)dx-\int_{x}^{+\infty}
\frac{\partial}{\partial x}\left(\widetilde{\phi}_{1}(x,\lambda)\widetilde{\psi}_{2,\lambda}(x,\lambda)-\widetilde{\psi}_{1,\lambda}(x,\lambda)
\widetilde{\phi}_{2}(x,\lambda)\right)dx.
\end{aligned}
\end{equation}
When $\lambda=\widetilde{\lambda}_{j}$, we note that $\widetilde{\phi}_{1}(x,\widetilde{\lambda}_{j})=\widetilde{b}_{j}\widetilde{\psi}_{1}(x,\widetilde{\lambda}_{j})$. Thus
\begin{equation}
\widetilde{a}_{\lambda}(\widetilde{\lambda}_{j})=\int_{-\infty}^{+\infty}2i\widetilde{b}_{j}
\widetilde{\psi}_{1}(x,\widetilde{\lambda}_{j})\widetilde{\psi}_{2}(x,\widetilde{\lambda}_{j})dx,
\end{equation}
that is
\begin{equation}
\int_{-\infty}^{+\infty}-2i\widetilde{c}_{j}\widetilde{\psi}_{1}(x,\widetilde{\lambda}_{j})
\widetilde{\psi}_{2}(x,\widetilde{\lambda}_{j})dx=1,
\end{equation}
where we define that $\widetilde{c}_{j}=-\frac{\widetilde{b}_{j}}{\widetilde{a}^{'}(\widetilde{\lambda}_{j})}$.\\
$\mathbf{Lemma ~3.5.}$ Suppose that $\widetilde{\psi}\left(x,t,\lambda\right)$ is a solution of spectral problem \eqref{lax1}, $M$ and $N$ satisfy the zero curvature equation \eqref{zero1}, then
\begin{equation}\label{lem3.2}
P\left(x,t,\lambda\right)=\widetilde{\psi}_{t}\left(x,t,\lambda\right)-N\widetilde{\psi}\left(x,t,\lambda\right)
\end{equation}
is a solution of spectral problem \eqref{lax1} as well.\\
$\mathbf{Theorem~ 3.2}$ The discrete scattering data ${\widetilde{\lambda}_{j}(t), \widetilde{c}_{j}(t)}$ for $\widetilde{\psi}(x,t,\lambda)$ and $\widetilde{\phi}(x,t,\lambda)$ posses the following time dependence. For $n=0$,
\begin{equation}
\begin{cases}
\widetilde{\lambda}_{j}(t)=e^{t+\ln \widetilde{\lambda}_{j}},\\
\widetilde{c}_j(t)=e^{t+\ln \widetilde{c}_j(t)},
\end{cases}
\end{equation}

For $n\geq 1$,
\begin{equation}
\begin{cases}
\widetilde{\lambda}_{j}(t)=\left(\widetilde{\lambda}_{j}(0)^{-n}-nt\right)^{-\frac{1}{n}},\\
\widetilde{c}_j(t)=\frac{\widetilde{c}_{j}(0)}{\widetilde{\lambda}_{j}(0)^{n+1}}\left(\widetilde{\lambda}_{j}(0)^{-n}-nt\right)^{-\frac{n+1}{n}},
\end{cases}
\end{equation}
where $\widetilde{\lambda}_j(0)$ and $\widetilde{c}_j(0)$ are the scattering data of spectral problem \eqref{lax1} in terms of $u(0, x)$ and $v(0, x)$.\\
$\mathbf{Proof}$. Based on Lemma 3.5, we have
\begin{equation}
\widetilde{\psi}_{t}(x,\widetilde{\lambda}_{j})-N\widetilde{\psi}(x,\widetilde{\lambda}_{j})=\gamma \widetilde{\psi}(x,\widetilde{\lambda}_{j}),
\end{equation}
where $\gamma$ is a constant to be determined here. Similar with \eqref{th3.12}, we have
\begin{equation}
\int_{-\infty}^{+\infty}\left(-(B+C)\widetilde{\psi}_{2}^{2}(x,\widetilde{\lambda}_{j})
-(B-C)\widetilde{\psi}_{1}^{2}(x,\widetilde{\lambda}_{j})\right)dx =\gamma.
\end{equation}
Therefore,
\begin{equation}
\begin{aligned}
\gamma &=-\int_{-\infty}^{+\infty}\left(\begin{array}{c}
B+C \\
B-C
\end{array}\right)\left(\begin{array}{cc}
\widetilde{\psi}_{2}(x,\widetilde{\lambda}_{j}) \\
\widetilde{\psi}_{1}(x,\widetilde{\lambda}_{j})
\end{array}\right)\\
&=-\sum_{k=0}^{n}\int_{-\infty}^{+\infty}\widetilde{\lambda}_{j}^{k}\left(\begin{array}{cc}
b_{1}+c_{0} \\
b_{1}-c_{0}
\end{array}\right)\left(\begin{array}{c}
\widetilde{\psi}_{2}^{2}(x,\widetilde{\lambda}_{j}) \\
\widetilde{\psi}_{1}^{2}(x,\widetilde{\lambda}_{j})
\end{array}\right)\widetilde{\lambda}_{j}^{n-k}dx\\
&=\frac{n+1}{2}\widetilde{\lambda}_{j}^{n}.
\end{aligned}
\end{equation}
Through \eqref{asym2}, it is noted that\\
$~~~~~~~~~~~~~~~~~~~~~~~~~~~~~~~~~~~~~~~~~~~~~
\widetilde{\psi}(x,\widetilde{\lambda}_{j})\sim \left(\begin{array}{c}
1 \\
0
\end{array}\right)e^{i\widetilde{\lambda}_{j} x} \sqrt{-i\widetilde{c}_{j}}$ as $x\rightarrow-\infty$,\\
which indicate that
\begin{equation}
\begin{aligned}
&\widetilde{\psi}_{t}(x,\widetilde{\lambda}_{j})-N\widetilde{\psi}(x,\widetilde{\lambda}_{j})\\
=&i\widetilde{\lambda}_{j,t}x\left(\begin{array}{c}
1 \\
0
\end{array}\right)e^{i\widetilde{\lambda}_{j}x}\sqrt{-i\widetilde{c}_{j}}+\left(\begin{array}{c}
1 \\
0
\end{array}\right)e^{i\widetilde{\lambda}_{j}x}\left(\sqrt{-i\tilde{c}_{j}}\right)_{t}\\
&-\left(\begin{array}{cc}
ix\widetilde{\lambda}_{j}^{n+1}  &  0 \\
0   &  -ix\widetilde{\lambda}_{j}^{n+1}
\end{array}\right)\left(\begin{array}{c}
1 \\
0
\end{array}\right)e^{i\widetilde{\lambda}_{j}x}\sqrt{-i\tilde{c}_{j}}\\
=&\frac{n+1}{2}\widetilde{\lambda}_{j}^{n}\left(\begin{array}{c}
1 \\
0
\end{array}\right)e^{i\widetilde{\lambda}_{j}x}\sqrt{-i\widetilde{c}_{j}}.
\end{aligned}
\end{equation}
Thus we obtain \\
$~~~~~~~~~~~~~~~~~~~~~~~~~~~~~~~~~~~~~~~~~~~~~~~~~~
\left(\sqrt{-i\tilde{c}_{j}}\right)_{t}=\frac{n+1}{2}\widetilde{\lambda}_{j}^{n}\sqrt{-i\tilde{c}_{j}}$\\
and\\
$~~~~~~~~~~~~~~~~~~~~~~~~~~~~~~~~~~~~~~~~~~~~~~~~~~~~~~
\tilde{c}_{j,t}=(n+1)\widetilde{\lambda}_{j}^{n}\tilde{c}_{j}.$\\

~~~~~~~~~~~~~~~~~~~~~~~~~~~~~~~~~~~~~~~~~~~~~~~~~~~~~~~~~~~~~~~~~~~~~~
~~~~~~~~~~~~~~~~~~~~~~~~~~~~~~~~~~~~~~~~~~~~~~~~~~~~~~~~~~~$\Box$

\subsection{General solutions for the non-isospectral TD hierarchy}

From \eqref{glm1} and \eqref{glm2} respectively,
\begin{equation}
\widetilde{\psi}(x,\widetilde{\lambda}_{k})=\left(\begin{array}{c}
1 \\
0
\end{array}\right)
e^{i\widetilde{\lambda}_{k}x+\int_{-\infty}^{x}udx}+\sum_{j=1}^{N}\frac{i}{\widetilde{\lambda}_{k}-\lambda_{j}}
c_{j}\psi(x,\lambda_{j})e^{i(\widetilde{\lambda}_{k}-\lambda_{j})x},
\end{equation}
and
\begin{equation}
\psi(x,\lambda_{k})=\left(\begin{array}{c}
0 \\
1
\end{array}\right)
e^{-i\lambda_{k}x-\int_{-\infty}^{x}udx}+\sum_{j=1}^{N}\frac{i}{\lambda_{k}-\widetilde{\lambda}_{j}}
\widetilde{c}_{j}\widetilde{\psi}(x,\widetilde{\lambda}_{j})e^{i(\widetilde{\lambda}_{j}-\lambda_{k})x}.
\end{equation}
Thus we have
\begin{equation}\label{solution1}
\widetilde{\psi}_{1}(x,\widetilde{\lambda}_{k})e^{-i\widetilde{\lambda}_{k}x-\int_{-\infty}^{x}udx}=1+
\sum_{j=1}^{N}\frac{i}{\widetilde{\lambda}_{k}-\lambda_{j}}
c_{j}\psi_{1}(x,\lambda_{j})e^{-i\widetilde{\lambda}_{j}x-\int_{-\infty}^{x}u dx},
\end{equation}
and
\begin{equation}\label{solution2}
\psi_{1}(x,\lambda_{k})e^{i\lambda_{k}x+\int_{-\infty}^{x}udx}=
\sum_{j=1}^{N}\frac{i}{\lambda_{k}-\widetilde{\lambda}_{j}}
\widetilde{c}_{j}\widetilde{\psi}_{1}(x,\widetilde{\lambda}_{j})e^{i\widetilde{\lambda}_{j}x+\int_{-\infty}^{x}u dx}.
\end{equation}
By \eqref{solution1} and \eqref{solution2}, we get
\begin{equation}\label{solution3}
\begin{aligned}
&~~~\psi_{1}(x,\lambda_{k})e^{-i\lambda_{k}x-\int_{-\infty}^{x}udx}\\
&=\sum_{j=1}^{N}\frac{i}{\lambda_{k}-\widetilde{\lambda}_{j}}
\widetilde{c}_{j}\widetilde{\psi}_{1}(x,\widetilde{\lambda}_{j})e^{i(\widetilde{\lambda}_{j}-2\lambda_{k})x-\int_{-\infty}^{x}u dx}\\
&=\sum_{j=1}^{N}\frac{i}{\lambda_{k}-\widetilde{\lambda}_{j}}
\widetilde{c}_{j}\widetilde{\psi}_{1}(x,\widetilde{\lambda}_{j})e^{-i\widetilde{\lambda}_{j}x-\int_{-\infty}^{x}udx}
\cdot e^{2i(\widetilde{\lambda}_{j}-\lambda_{k})x}\\
&=\sum_{j=1}^{N}\frac{i}{\lambda_{k}-\widetilde{\lambda}_{j}}\widetilde{c}_{j}e^{2i(\widetilde{\lambda}_{j}-\lambda_{k})x}
\cdot \left(1+\sum_{k=1}^{N}\frac{i}{\widetilde{\lambda}_{j}-\lambda_{k}}
c_{k}\psi_{1}(x,\lambda_{k})e^{-i\widetilde{\lambda}_{k}x-\int_{-\infty}^{x}u dx} \right),
\end{aligned}
\end{equation}

Then we multiply the left and right sides of \eqref{solution3} by $c_{k}$, and get
\begin{equation}
\begin{aligned}\label{solution4}
&~~~c_{k}\psi_{1}(x,\lambda_{k})e^{-i\lambda_{k}x-\int_{-\infty}^{x}udx}
-c_{k}\sum_{j=1}^{N}\frac{i}{\lambda_{k}-\widetilde{\lambda}_{j}}\widetilde{c}_{j}
\cdot \left(\sum_{k=1}^{N}\frac{i}{\widetilde{\lambda}_{j}-\lambda_{k}}
c_{k}\psi_{1}(x,\lambda_{k})e^{-i\widetilde{\lambda}_{k}x-\int_{-\infty}^{x}u dx}\right)e^{2i(\widetilde{\lambda}_{j}-\lambda_{k})x}\\
&=c_{k}\sum_{j=1}^{N}\frac{i}{\lambda_{k}-\widetilde{\lambda}_{j}}\widetilde{c}_{j}
e^{2i(\widetilde{\lambda}_{j}-\lambda_{k})x}.
\end{aligned}
\end{equation}
This is a solvable system of $N$-member linear equations with respect to $c_{k}\psi_{1}(x,\lambda_{k})e^{-i\lambda_{k}x-\int_{-\infty}^{x}udx}$.

$\mathbf{Theorem~ 3.3}$
\begin{equation}\label{v}
\frac{iv^2}{2}=\sum_{j=1}^{N}ic_{j}\left(\psi_{1}(x,\lambda_{j})e^{-i\lambda_{j}x-\int_{-\infty}^{x}udx}\right)_{x}.
\end{equation}
$\mathbf{Proof.}$
Since \eqref{psy1}, we know $\widetilde{\psi}_{1}(x,\lambda)e^{-i\lambda x-\int_{-\infty}^{x}udx+o(\lambda^{-2})}$ has the following asymptotic property that
\begin{equation}\label{v1}
\begin{aligned}
&~~~\widetilde{\psi}_{1}(x,\lambda)e^{-i\lambda x-\int_{-\infty}^{x}udx+o(\lambda^{-2})}\\
&=e^{\int_{-\infty}^{x}-\frac{iv^2}{2\lambda}dx+o(\lambda^{-2})},\\
&=1+\frac{1}{2\lambda}\int_{-\infty}^{x}iv^2dx+o(\lambda^{-2})
\end{aligned}
\end{equation}
as $\lambda\rightarrow\infty$.

Thus we can obtain an expression for $v$ by deriving both sides of \eqref{v1} with respect to $x$,
\begin{equation}\label{v2}
\begin{aligned}
\frac{iv^2}{2}=\lim_{\lambda\rightarrow\infty} \lambda \left(\widetilde{\psi}_{1}(x,\lambda)e^{-i\lambda x-\int_{-\infty}^{x}udx} -1\right)_x.
\end{aligned}
\end{equation}
Substituting \eqref{glm1} into \eqref{v2} to replace $\widetilde{\psi}_{1}(x,\lambda)e^{-i\lambda x-\int_{-\infty}^{x}udx}$ with $\psi_{1}(x,\lambda_{j})e^{-i\lambda_{j}x-\int_{-\infty}^{x}udx}$, we obtain

\begin{equation}\label{v3}
\begin{aligned}
\frac{iv^2}{2}=\lim_{\lambda\rightarrow\infty} \sum_{j=1}^{N}\frac{i\lambda}{\lambda-\lambda_{j}} \left(c_{j}\psi_{1}(x,\lambda_{j})e^{-i\lambda_{j}x-\int_{-\infty}^{x}u dx}\right)_x.
\end{aligned}
\end{equation}
Thus the Theorem 3.3 is proved.
~~~~~~~~~~~~~~~~~~~~~~~~~~~~~~~~~~~~~~~~~~~~~~~~~~~~~~~~~~~~~~~~~~~~~~~~~~$\Box$

From the solvable system \eqref{solution4} and Theorem 3.3, we can derive that
\begin{equation}\label{vns}
v=\pm \sqrt{2\left((D^{-1}F)^{T}\cdot \left(\begin{array}{c}
1 \\
\vdots \\
1
\end{array}\right)\right)_{x}~~},
\end{equation}
where\\
 $~~~~~~~~~~~~~~~~~~~~~~~~~~~~~~~~~~~~~~~~~~~~~~~~~~~~~~~~~~~~D=I+d_{N\times N}$\\
 with\\
 $~~~~~~~~~~~~~~~~~~~~~~~~~~~~~~~~~~~~~~~~~~~~~~~~~~~~
 d_{kj}=c_{k}\sum_{j=1}^{N}\frac{\widetilde{c}_{j}e^{2i(\widetilde{\lambda}_{j}-\lambda_{j})x}}
 {\left(\lambda_{k}-\widetilde{\lambda}_{j}\right)\left(\widetilde{\lambda}_{j}-\lambda_{k}\right)}$,\\
 and \\
 $~~~~~~~~~~~~~~~~~~~~~~~~~~~~~~~~~~~~~~~~~~~~~~~~~~~~~~~~~~~~F=\left(\begin{array}{c}
\beta_{1} \\
\vdots \\
\beta_{N}
\end{array}\right)$\\
with\\
$~~~~~~~~~~~~~~~~~~~~~~~~~~~~~~~~~~~~~~~~~~~~~~
\beta_{k}=c_{k}\sum_{k=1}^{N}\frac{i}{\lambda_{k}-\widetilde{\lambda}_{k}}\widetilde{c}_{k}
e^{2i(\widetilde{\lambda}_{k}-\lambda_{k})x}$.

By now we have obtained an explicit expression for the $N$-soliton solution of $v$. It is worth noting that the expression for $u$ can be obtained by substituting $v$ in the second equation of \eqref{equ1}. In this case, the second equation of \eqref{equ1} is an ordinary differential equation with respect to $u$ as $v=$\eqref{vns}.

\section{l-soliton-like solution}
In this section, we follow the example of the previous section to obtain the explicit expression for the l-soliton-like solution of non-isospectral TD hierarchy as $N = 1$.

Let $\lambda=\widetilde{\lambda}_{1}$ in \eqref{glm1}, we note that
\begin{equation}\label{s2}
\widetilde{\psi}_{1}(x,\widetilde{\lambda}_{1})e^{-i\widetilde{\lambda}_{1} x-\int_{-\infty}^{x}u dx}=1+\frac{i}{\widetilde{\lambda}_{1}-\lambda_{1}}c_{1}\psi_{1}(x,\lambda_{1})
e^{-i\lambda_{1}x-\int_{-\infty}^{x}udx}.
\end{equation}
Similarly in \eqref{glm2}, let $\lambda=\lambda_{1}$,we get
From \eqref{glm2}, we get
\begin{equation}\label{s3}
e^{i\lambda_{1}x+\int_{-\infty}^{x}udx}\psi_{1}(x,\lambda_{1})=\frac{i}{\lambda_{1}-\widetilde{\lambda}_{1}}\widetilde{c}_{1}
\widetilde{\psi}_{1}(x,\widetilde{\lambda}_{1})e^{i\widetilde{\lambda}_{1}x+\int_{-\infty}^{x}udx}.
\end{equation}
Substituting \eqref{s3} into \eqref{s2} yields
\begin{equation}
\begin{cases}
\widetilde{\psi}_{1}(x,\widetilde{\lambda}_{1})e^{-i\widetilde{\lambda}_{1}x-\int_{-\infty}^{x}udx}=
1+\frac{i}{\widetilde{\lambda}_{1}-\lambda_{1}}c_{1}\widetilde{c}_{1}\frac{i}{\lambda_{1}-\widetilde{\lambda}_{1}}
\widetilde{\psi}_{1}(x,\widetilde{\lambda}_{1})e^{i\widetilde{\lambda}_{1}x+\int_{-\infty}^{x}udx}
e^{-2i\lambda_{1}x-2\int_{-\infty}^{x}udx},\\
\psi_{1}(x,\lambda_{1})e^{i\lambda_{1}x+\int_{-\infty}^{x}udx}=\frac{i}{\lambda_{1}-\widetilde{\lambda}_{1}}\widetilde{c}_{1}
e^{2i\widetilde{\lambda}_{1}x+2\int_{-\infty}^{x}udx}\left(1+\frac{i}{\widetilde{\lambda}_{1}-\lambda_{1}}c_{1}\psi_{1}(x,\lambda_{1})
e^{-i\lambda_{1}x-\int_{-\infty}^{x}udx}\right).
\end{cases}
\end{equation}
Let\\
 $~~~~~~~~~~~~~~~~~~~~~~~~~~~~~~~~~~~~~~~~~~~~~~~~~~~~~~~~~
 T=\psi_{1}(x,\lambda_{j})e^{-i\lambda_{j}x-\int_{-\infty}^{x}udx}$, \\
 we have\\
$~~~~~~~~~~~~~~~~~~~~~~~~~~~~~~~~~~~~~~~~~~~~~~~~~
T=\frac{i}{\lambda_{1}-\widetilde{\lambda}_{1}}\widetilde{c}_{1}e^{2i(\widetilde{\lambda}_{1}-\lambda_{1})x}\left(1+\frac{i}{\widetilde{\lambda}_{1}
-\lambda_{1}}c_{1}T\right)$.\\
So that
\begin{equation}
\begin{aligned}
T&=\frac{\frac{i}{\lambda_{1}-\widetilde{\lambda}_{1}}\widetilde{c}_{1}e^{2i(\widetilde{\lambda}_{1}-\lambda_{1})x}}
{1-\frac{1}{(\lambda_{1}-\widetilde{\lambda}_{1})^{2}}c_{1}\widetilde{c}_{1}e^{2i(\widetilde{\lambda}_{1}-\lambda_{1})x}}\\
&=\frac{i}{\lambda_{1}-\widetilde{\lambda}_{1}}\widetilde{c}_{1}\cdot\frac{1}{e^{2i(\lambda_{1}-\widetilde{\lambda}_{1})x}
-\frac{1}{(\lambda_{1}-\widetilde{\lambda}_{1})^{2}}c_{1}\widetilde{c}_{1}}.
\end{aligned}
\end{equation}
Therefore, we get
\begin{equation}
\begin{aligned}
 T_{x}&=\frac{i}{\lambda_{1}-\widetilde{\lambda}_{1}} \tilde{c}_{1} \cdot \frac{-e^{2 i\left(\lambda_{1}-\widetilde{\lambda}_{1}\right) x} \cdot 2 i\left(\lambda_{1}-\widetilde{\lambda}_{1}\right)}{\left(e^{2 i\left(\lambda_{1}-\widetilde{\lambda}_{1}\right) x}-\frac{1}{\left(\lambda_{1}-\widetilde{\lambda}_{1}\right)^{2}} c_{1} \widetilde{c}_{1}\right)^{2}} \\
&=\frac{2 \widetilde{c}_{1} e^{2 i\left(\lambda_{1}-\widetilde{\lambda}_{1}\right) x}}{\left(e^{2 i\left(\lambda_{1}-\widetilde{\lambda}_{1}\right) x}-\frac{1}{\left(\lambda_{1}-\widetilde{\lambda}_{1}\right)^{2}} c_{1} \widetilde{c}_{1}\right)^{2}} \\
&=\frac{2 \widetilde{c}_{1}}{\left(e^{i\left(\lambda_{1}-\widetilde{\lambda}_{1}\right) x}-\frac{1}{\left(\lambda_{1}-\widetilde{\lambda}_{1}\right)^{2}} c_{1} \widetilde{c_{1}} e^{i\left(\widetilde{\lambda}_{1}-\lambda_{1}\right) x}\right)^{2}} \\
&=\frac{2 \widetilde{c}_{1} \frac{\left(\lambda_{1}-\widetilde{\lambda}_{1}\right)^{2}}{c_{1} \widetilde{c}_{1}}}{\left(\frac{\lambda_{1}-\widetilde{\lambda}_{1}}{\sqrt{c_{1} \tilde{c}_{1}}} e^{i\left(\lambda_{1}-\widetilde{\lambda}_{1}\right) x}-\frac{\sqrt{c_{1} \widetilde{c}_{1}}}{\lambda_{1}-\widetilde{\lambda}_{1}} e^{i\left(\widetilde{\lambda}_{1}-\lambda_{1}\right) x}\right)^{2}}.
\end{aligned}
\end{equation}
Since $csch x=\frac{2}{e^{x}-e^{-x}}$, then we can derive that
\begin{equation}
T_{x}=\frac{(\lambda_{1}-\widetilde{\lambda}_{1})^{2} csch^{2}\left(i(\lambda_{1}-\widetilde{\lambda}_{1})x
+\ln \left(\frac{\lambda_{1}-\widetilde{\lambda}_{1}}{\sqrt{c_{1}\widetilde{c}_{1}}} \right) \right) }{c_{1}}.
\end{equation}
Finally, we get the explicit expression for $v$ that
\begin{equation}
\begin{aligned}
v^{2}&=2c_{1}T_{x}\\
&=2(\lambda_{1}-\widetilde{\lambda}_{1})^{2} csch^{2}\left(i(\lambda_{1}-\widetilde{\lambda}_{1})x
+\ln \left(\frac{\lambda_{1}-\widetilde{\lambda}_{1}}{\sqrt{c_{1}\widetilde{c}_{1}}} \right) \right),
\end{aligned}
\end{equation}
that is
\begin{equation}\label{vs}
v=\pm \sqrt{2}(\lambda_{1}-\widetilde{\lambda}_{1})^{2} csch\left(i(\lambda_{1}-\widetilde{\lambda}_{1})x
+\ln \left(\frac{\lambda_{1}-\widetilde{\lambda}_{1}}{\sqrt{c_{1}\widetilde{c}_{1}}} \right) \right).
\end{equation}

After a simplification operation, we find that the expression for the solution \eqref{vs} can be represented by a hyperbolic function, which is similar to the soliton solution of an isospectral equation. In addition, we present the figure of l-soliton-like solution for the second nonisospectral equation in the non-isospectral TD equation \eqref{equ2}.

As shown in Fig. 3, the figure of the 1-soliton-like solution also has the similar properties with the 1-soliton solution of the isospectral equation. The solution $|v|$ increases with $x$ before $|v|$ reaches its maximum, and decreases with $x$ after $|v|$ reaches its maximum. However, the maximum value of 1-soliton solutions of isospectral equations remain unchange with time while the maximum value of 1-soliton-like solution evolves with time.

\begin{figure}
  \centering
  \includegraphics[width=5cm]{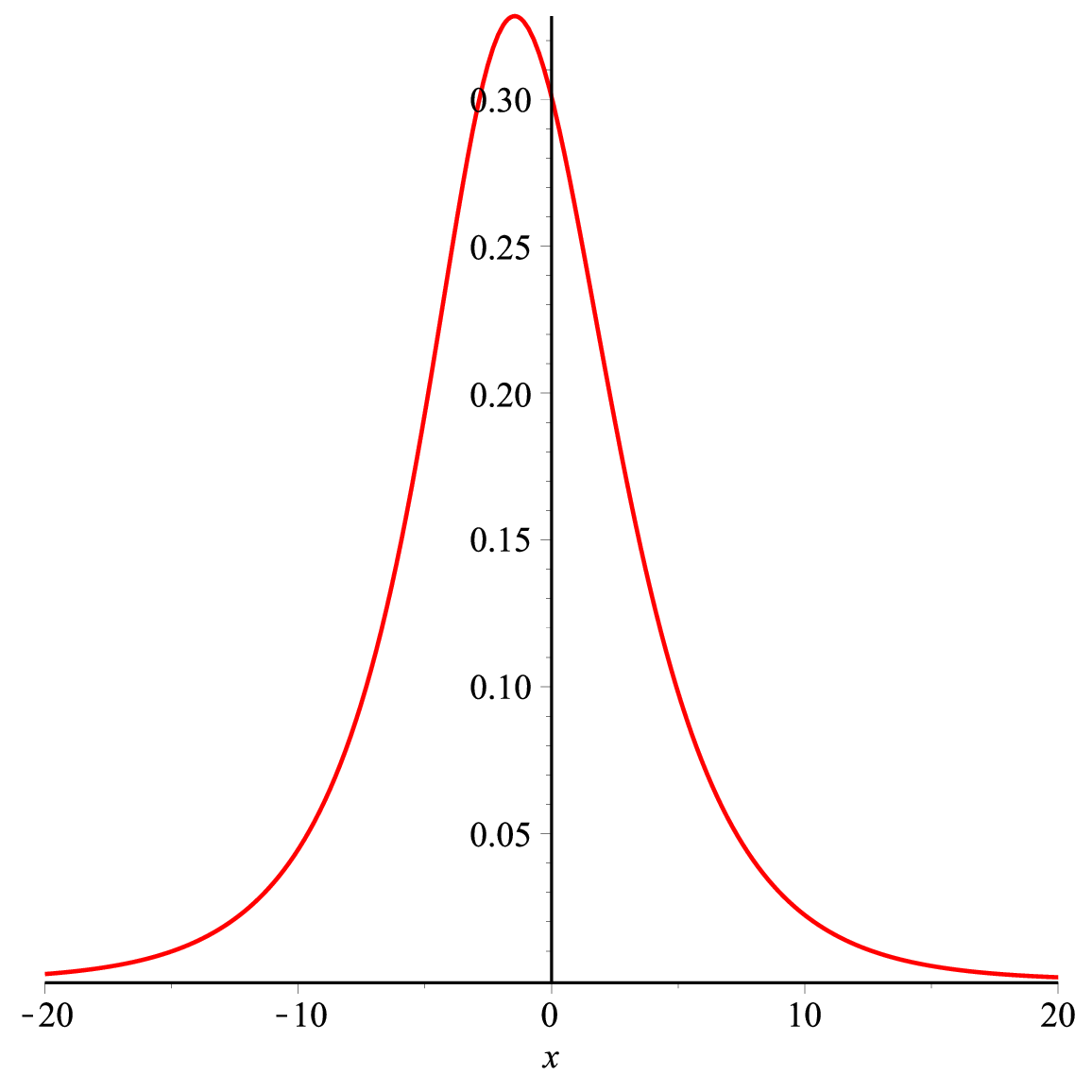}\includegraphics[width=5cm]{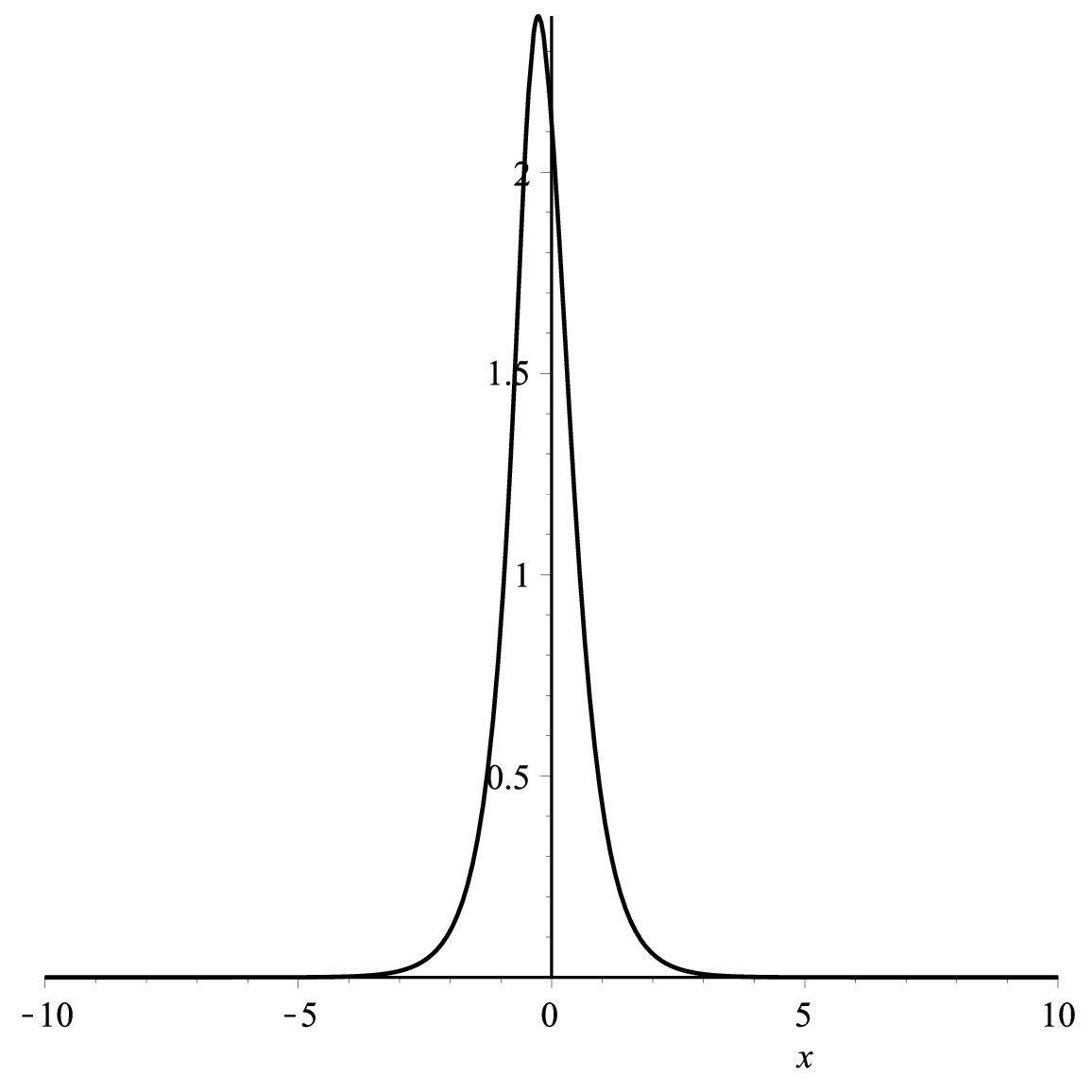}\includegraphics[width=5cm]{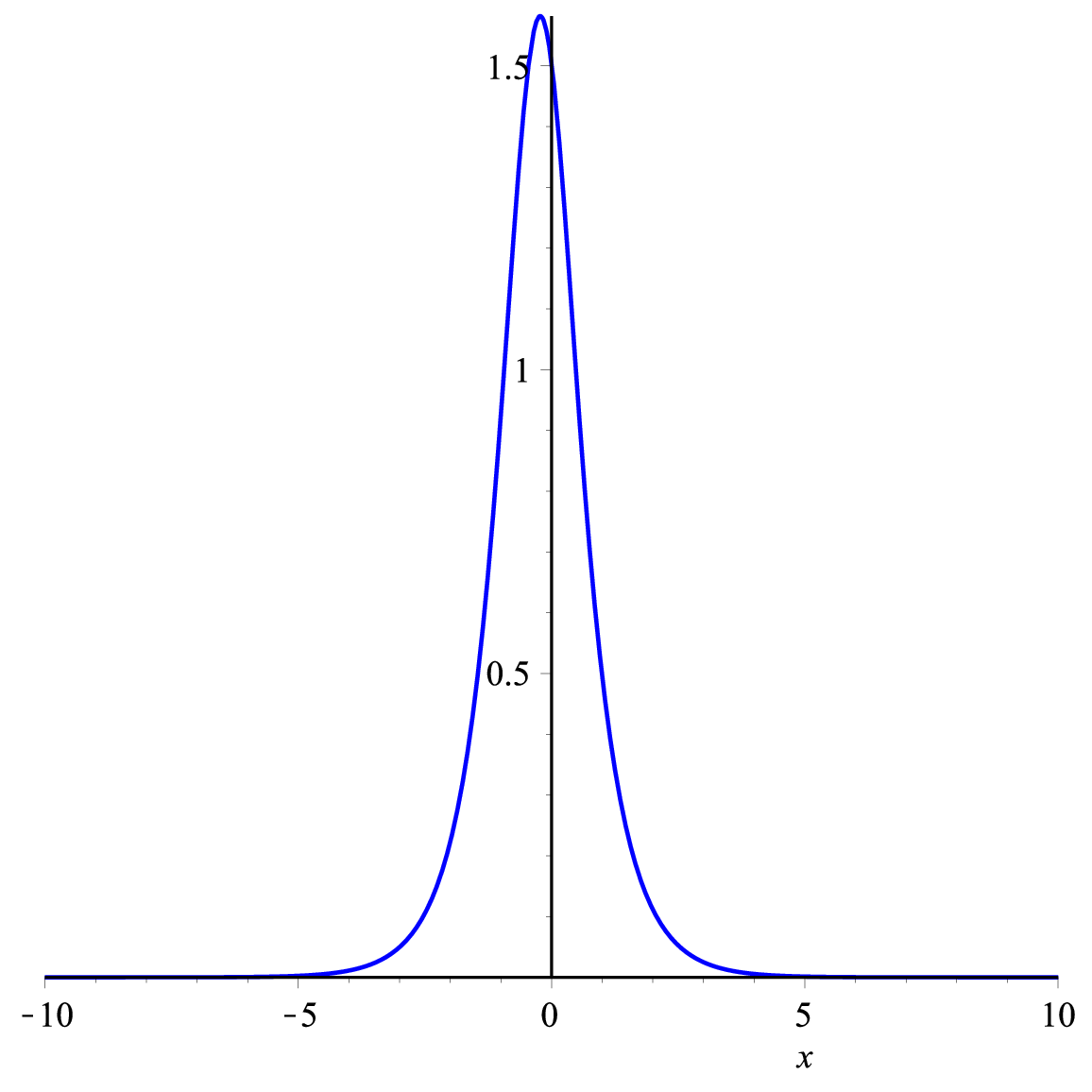}
  \caption{1-soliton-like solution as n=1. $\lambda_1(0)=1+i$, $\widetilde{\lambda}_1(0)=2-i$, $c_1(0)=\lambda_1(0)^{2}$, $\widetilde{c}_1(0)=\widetilde{\lambda}_1(0)^{2}$. (a): $t=-1$; (b): $t=0$; (c): $t=1$;}
\end{figure}

\section{Conclusion}
In this paper, we introduced a systematic approach to solve the initial value problem associated with the non-isospectral TD hierarchy through the utilization of the inverse scattering transform. Given our consideration of the non-isospectral case, we acquired the time evolution of the scattering data, which is different from that of the corresponding isospectral TD hierarchy. Finally, we obtained the exact solutions of this hierarchy under various parameters. Moreover, these exact solutions can also be verified by substituting them into the corresponding equations.
There are several problems to be considered in the future. How to research rogue waves of equations presented in the paper in terms of the approach \cite{Zhang19,Zhang191}.

\section*{Acknowledgements}

This work was supported by the National Natural Science Foundation of China grant No.12371256;
 the National Natural Science Foundation of China grant No.11971475.\\

\section*{Data Availability}
Data sharing is not applicable to this article as no new data were created or analyzed in this study.


\end{CJK*}
\end{document}